\documentclass[sigconf]{acmart} 
\AtBeginDocument{%
  \providecommand\BibTeX{{%
    \normalfont B\kern-0.5em{\scshape i\kern-0.25em b}\kern-0.8em\TeX}}}

\copyrightyear{2024}
\acmYear{2024}
\setcopyright{acmlicensed}\acmConference[CHI '24]{Proceedings of the CHI Conference on Human Factors in Computing Systems}{May 11--16, 2024}{Honolulu, HI, USA}
\acmBooktitle{Proceedings of the CHI Conference on Human Factors in Computing Systems (CHI '24), May 11--16, 2024, Honolulu, HI, USA}
\acmDOI{10.1145/3613904.3642292}
\acmISBN{979-8-4007-0330-0/24/05}

\usepackage{subcaption}
\usepackage[acronym,xindy,toc,numberedsection]{glossaries}
            \newglossary[nlg]{notation}{not}{ntn}{Notation} 	
            \newacronym{alsep}{ALSEP}{Apollo Lunar Surface Experiment Package}
\newacronym{aisep}{AISEP}{Apollo Inspired Surface Experiment Package}
\newacronym{anova}{ANOVA}{Analysis of Variances}
\newacronym{api}{API}{Application Programming Interface}
\newacronym{ar}{AR}{Augmented Reality}
\newacronym{ascan}{AsCan}{Astronaut Candidate}
\newacronym{basalt}{BASALT}{Biological Analog Science Associated with Lava Terrains}
\newacronym{cad}{CAD}{Computer-Assisted Design}
\newacronym{ccig}{CCIG}{Cold Cathode Ion Gage}
\newacronym{conops}{ConOps}{Concepts of Operations}
\newacronym{cpu}{CPU}{Core Processing unit}
\newacronym{drt}{DRT}{Dome Removal Tool}
\newacronym{dtm}{DTM}{Digital Terrain Model}
\newacronym{eac}{EAC}{European Astronaut Center}
\newacronym{easep}{EASEP}{Early Apollo Surface Experiment Package}
\newacronym{el3}{EL3}{European Large Logistics Lander}
\newacronym{esa}{ESA}{European Space Agency}
\newacronym{eurocom}{EUROCOM}{European Communicator and Medical Operations}
\newacronym{eva}{EVA}{Extravehicular Activity}
\newacronym{expert}{ExPeRT}{Exploration Preparation, Research and Technology Team}
\newacronym{fov}{FOV}{Field of View}
\newacronym{ftt}{FTT}{Fuel Transfer Tool}
\newacronym{gpu}{GPU}{Graphics Processing Unit}
\newacronym{hci}{HCI}{Human-Computer Interactions}
\newacronym{hitl}{HITL}{Human-In-The-Loop}
\newacronym{hmd}{HMD}{Head-Mounted Display}
\newacronym{ipq}{IPQ}{IGroup Presence Questionnaire}
\newacronym{iss}{ISS}{International Space Station}
\newacronym{ive}{IVE}{Immersive Virtual Environment}
\newacronym{jpl}{JPL}{Jet Propulsion Laboratory}
\newacronym{jsc}{JSC}{Johnson Space Center}
\newacronym{lec}{LEC}{Lunar Equipement Conveyor}
\newacronym{lem}{LEM}{Lunar Excursion Module}
\newacronym{lod}{LOD}{Level Of Details}
\newacronym{lroc}{LROC}{Lunar Reconnaissance Orbiter Camera}
\newacronym{lsm}{LSM}{Lunar Surface Magnetometer}
\newacronym{mr}{MR}{Mixed Reality}
\newacronym{nasa}{NASA}{National Aeronautics and Space Administration}
\newacronym{nbf}{NBF}{Neutral Buoyancy Facility}
\newacronym{pse}{PSE}{Passive Seismic Experiment}
\newacronym{qec}{QEC}{Quick Exposure Check}
\newacronym{reba}{REBA}{Rapid Entire Body Assessment}
\newacronym{riot}{RIOT}{Rapid Integration and Organisation of Tasks}
\newacronym{rtg}{RTG}{Radioisotope Thermoelectric Generator}
\newacronym{side}{SIDE}{Suprathermal Ion Detector Experiment}
\newacronym{sme}{SME}{Subject Matter Expert}
\newacronym{sus}{SUS}{System Usability Scale}
\newacronym{sws}{SWS}{Solar Wind Spectrometer}
\newacronym{tlx}{TLX}{Task Load Index}
\newacronym{rtlx}{RTLX}{Raw Task Load Index}
\newacronym{ue4}{UE4}{Unreal Engine 4}
\newacronym{uht}{UHT}{Universal Handling Tool}
\newacronym{ui}{UI}{User Interface}
\newacronym{veq}{VEQ}{Virtual Embodiment Questionnaire}
\newacronym{vr}{VR}{Virtual Reality}
\newacronym{vrastra12}{VR-ASTRA12}{Virtual Reality to Assess Scenarios Through Recreating Apollo}
\newacronym{wiibb}{WiiBB}{Wii Blance Board}
\newacronym{xemu}{xEMU}{exploration Extravehicular Mobility Unit}
\newacronym{xr}{XR}{eXtended Reality}
\newacronym{xrlab}{XR Lab}{Extended Reality Laboratory}

\begin{document}

\title[Touching the Moon]{Touching the Moon: Leveraging Passive Haptics, Embodiment and Presence for Operational Assessments in Virtual Reality}

\author{Florian Dufresne}
\email{florian.dufresne@ensam.eu}
\authornote{Both authors contributed equally to this research.}
\affiliation{%
  \institution{Arts et Métiers Institute of Technology}
  \city{Changé}
  \country{France}
  \postcode{53000}
  }
\orcid{0000-0002-4664-8828}

\author{Tommy Nilsson}
\email{tommy.nilsson@esa.int}
\authornotemark[1]
\affiliation{%
  \institution{European Space Agency}
  \streetaddress{-}
  \city{Cologne}
  \country{Germany}
  \postcode{-}
}
\orcid{0000-0002-8568-0062}

\author{Geoffrey Gorisse}
\email{geoffrey.gorisse@ensam.eu}
\affiliation{%
  \institution{Arts et Métiers Institute of Technology}
  \city{Changé}
  \country{France}
  \postcode{53000}
  }
\orcid{0000-0003-1613-927X}

\author{Enrico Guerra}
\affiliation{%
  \institution{University Duisburg-Essen}
  \city{Duisburg}
  \country{Germany}}
\orcid{0000-0003-4531-6116}
\email{enrico.guerra@uni-due.de}

\author{André Zenner}
\email{andre.zenner@uni-saarland.de}
\orcid{0000-0003-3386-1635}
\affiliation{%
  \institution{Saarland University \& DFKI}
  \streetaddress{Saarland Informatics Campus}
  \city{Saarbrücken}
  \country{Germany}
}
\orcid{0000-0003-3386-1635}

\author{Olivier Christmann}
\email{olivier.christmann@ensam.eu}
\affiliation{%
  \institution{Arts et Métiers Institute of Technology}
  \city{Changé}
  \country{France}
  \postcode{53000}
  }
\orcid{0000-0001-8652-5630}

\author{Leonie Bensch}
\email{leonie.bensch@dlr.de}
\affiliation{%
  \institution{Institute for Software Technology - Software for Space Systems and Interactive Visualization}
  \city{Cologne}
  \country{Germany}
  }
\orcid{0000-0003-4736-5579}

\author{Nikolai Anton Callus}
\email{nikolaiantoncallus@gmail.com}
\affiliation{%
  \institution{European Space Agency}
  \streetaddress{-}
  \city{Cologne}
  \country{Germany}
  \postcode{-}
}
\orcid{0009-0008-3063-2697	}

\author{Aidan Cowley}
\email{aidan.cowley@esa.int}
\affiliation{%
  \institution{European Space Agency}
  \streetaddress{-}
  \city{Cologne}
  \country{Germany}
  \postcode{-}
}
\orcid{0000-0001-8692-6207}

\renewcommand{\shortauthors}{Dufresne and Nilsson, et al.}

\begin{abstract}
  Space agencies are in the process of drawing up carefully thought-out Concepts of Operations (ConOps) for future human missions on the Moon. These are typically assessed and validated through costly and logistically demanding analogue field studies. While interactive simulations in Virtual Reality (VR) offer a comparatively cost-effective alternative, they have faced criticism for lacking the fidelity of real-world deployments. This paper explores the applicability of passive haptic interfaces in bridging the gap between simulated and real-world ConOps assessments. Leveraging passive haptic props (equipment mockup and astronaut gloves), we virtually recreated the Apollo 12 mission procedure and assessed it with experienced astronauts and other space experts. Quantitative and qualitative findings indicate that haptics increased presence and embodiment, thus improving perceived simulation fidelity and validity of user reflections. We conclude by discussing the potential role of passive haptic modalities in facilitating early-stage ConOps assessments for human endeavours on the Moon and beyond.
\end{abstract}



\begin{teaserfigure}
  \includegraphics[width=\textwidth]{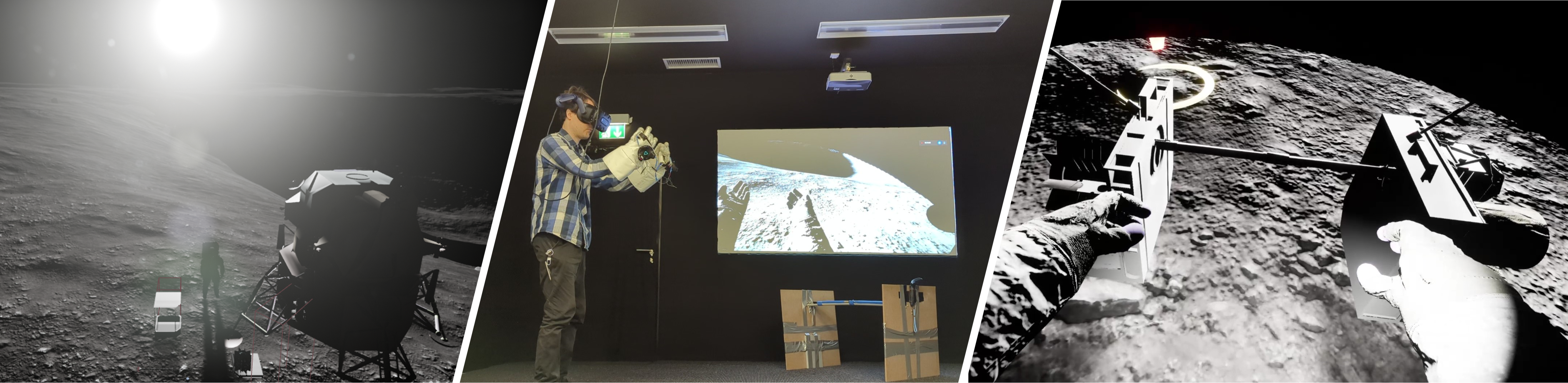}
  \caption{Visuals of the 3D environment created for the study presented in this paper. From left to right: overview of the surveyor 3 crater - view of an immersed participant wearing the astronaut gloves in front of the physical mockup - virtual representation of the assembled package (barbell).}
  \Description{Recreation of the Apollo 12 mission landing site}
  \label{fig:teaser}
  \Description{A figure constituted of three aligned pictures: a screenshot of the lunar virtual environment, a participant wearing a VR headset and in presence of the haptic props, and a screenshot from the immersed user's view.}
\end{teaserfigure}


\maketitle

\section{Introduction}\label{section: introduction}

50 years after astronauts last walked on the lunar surface, a renewed international effort centred around the \gls*{nasa}’s Artemis program is now setting the stage for the next chapter in humanity’s exploration of the Moon \cite{Coan2020Conops}. This endeavour will be underpinned by a wide spectrum of novel technological solutions ranging from \gls*{eva} suits to lunar landers and surface habitation modules. All such solutions will have to be integrated within carefully planned \gls*{conops}, defined by \citet{Beaton2019UsingExploration} as "\textit{the instantiation of operational design elements that guide the organisation and flow of personnel, communications, hardware, software, and data products involved in a mission concept}". 

Lunar \gls*{conops} will inherently entail considerable amount of manual work procedures  (e.g., payload deployment and maintenance of surface infrastructure) which will need to be assessed and fine-tuned ahead of the coming lunar expeditions. However, assessing such prospective lunar \gls*{conops} is not trivial. The lunar environment poses severe challenges, including blinding lights, pitch-black shadows, and reduced gravity. Future astronauts will have to navigate these obstacles carrying a heavy physical and mental workload, all while wearing bulky and restrictive spacesuits \cite{Vaniman1991LunarEnvironment}. \gls*{conops} assessments and crew training activities undertaken by space agencies have traditionally sought to replicate some of these challenges through the use of \textit{analogues} in the form of either natural terrains (e.g., large caves in Sardinia, Italy \cite{sauro2021}) or artificial facilities (e.g., \gls*{nbf} \cite{Bessone2015Analogues}). This practice is notorious for incurring substantial logistical costs, frequent project delays and budget overruns \cite{casini2020}. Analogue studies are likewise characterised by several limitations impairing their fidelity and utility, including the inability to realistically simulate the reduced gravity or the distinct lighting conditions found on the Moon. To circumvent these limitations, researchers are increasingly turning to alternative technologies, such as \gls*{vr} systems, to more efficiently simulate the unique lunar surface conditions and facilitate relevant \gls*{conops} evaluations \cite{Nilsson2023ShapeHumanity}.

A significant advantage of VR simulations is that they can be produced and used at a fraction of the time and cost of their real-world counterparts \cite{Banerjee2021AlternativeReality}. This makes VR an ideal tool for visualising lunar environmental conditions and interactively simulating prospective surface \gls*{conops} scenarios. Nevertheless, the predominantly audiovisual nature of most \gls*{vr} simulations and the consequent lack of fidelity compared to real-world analogue studies \cite{SilvaMartinez2021HumanInTheLoop} has attracted criticism questioning the validity of certain aspects of simulated lunar scenarios (e.g. the absence of simulated movement constraints \cite{Nilsson2022EL3}). 

Past research has shown that presence and the sense of embodiment constitute critical contributors to the fidelity of \gls*{vr} simulations~\cite{Jerald:2015:VRBook}. Such psychological states can be leveraged by adding accurate kinesthetic and tactile sensations, unified under the umbrella term of \textit{haptics}, through the use of proxies. Haptics have indeed proved to positively impact the sense of presence in an \gls*{ive}~\cite{Insko:2001:PHF} and the sense of embodiment towards the virtual body~\cite{Kilteni2012Embodiment, Froehner:2019:HapticsIncreaseEmbodiment}. This raises, then, an important question about the use of haptics for VR-based \gls*{conops} reviews: Do these findings about the advantages of passive haptics apply in the context of ConOps assessments? If so, what would consequently be their impact on the validity of VR-based operational evaluations?  

In an effort to tackle this question, we produced a \gls*{vr} reconstruction of a renowned and well documented \gls*{eva} in the form of the \gls*{alsep} deployment carried out during the Apollo 12 mission, which launched on the $14^{th} of November 1969$. This provided us with the basis for a side-by-side comparative study featuring 27 highly qualified human spaceflight experts, including experienced astronauts, instructors, engineers, managers, scientists, as well as a number of relevant students. Following a mixed-design approach, half of our participants completed the simulated mission using an Apollo 12-inspired physical mockup, and the other half only used its virtual counterpart, as a between-subjects variable. In addition, all participants completed the scenario twice: once while wearing authentic astronaut gloves, and once without them as a within-subjects variable. In addition to presence and embodiment, perceived task load concerning the pertinent \gls*{eva} interactions was measured. Qualitative reflections were likewise collected and subsequently compared to feedback elements provided by the actual Apollo 12 crew. Finally, the usability of our systems was measured to evaluate the acceptance of the passive haptic interfaces.

We hypothesised that wearing physical gloves would bring a tactile and kinesthetic sensation coherent with the virtual representations of the hands, which, in turn, would enhance the sense of embodiment toward the virtual body. We also expected that the gloves, the physical mockup, or both simultaneously, would increase the sense of presence in the virtual environment, but also general workload because of the operational constraints they introduce (a notable theme in the feedback provided by the Apollo 12 crew). In exploring these hypotheses and undertaking the work outlined above, we offer the following contributions:
\begin{itemize}
    \item an original system implementation tailored for the \gls*{conops} assessment of the Apollo 12 mission, including the props (physical mockup and gloves), along with a hand controller design that allows for seamless interaction with these props.
    \item a detailed quantitative comparison between a wearable and a handheld passive haptic interface delivering tactile and kinesthetic feedback, namely wearable gloves and a physical weighted barbell mockup. 
    \item a qualitative analysis of the VR configurations using different haptic interfaces in the context of \gls*{conops} assessment, performed by a panel of highly qualified experts. 
    \item a unique one-to-one comparison between the feedback provided by actual Apollo 12 astronauts and the feedback from human spaceflight experts who performed the mission's tasks virtually.
\end{itemize}
Our results show that the passive haptic interfaces increase the perceived sense of presence (gloves and combination of both proxies) and embodiment toward the virtual body (gloves), thus improving the overall fidelity of the simulation. The qualitative analysis further corroborates these findings, highlighting the important role of haptic feedback in assessing key aspects of lunar surface activities. Finally, comparing this feedback to that of the Apollo crew allows us to take a first step towards understanding the validity of VR-based ConOps assessments.

\section{Related work}\label{section: related work}

\subsection{Real-World ConOps Assessments}

ConOps are concerned with organising human resources, hardware and software into coherent operations while managing potential frictions between all these aspects. A well established approach to assessing such operational concepts relies on the use of so-called \textit{analogue} missions, i.e., field or lab studies in which efforts are made to accurately recreate certain aspects of planned missions. In this context, the \gls*{basalt} missions aimed at identifying the best instruments, techniques, training requirements and execution strategies in preparation of extraterrestrial field activities, can be mentioned as a notable example\cite{Beaton2020Basalt, Beaton2019Assessing, Beaton2019UsingExploration}. \citet{Beaton2019Assessing}, who took part in this endeavour, relied on the work of \citet{Abercromby2013EvaluationExploration} to assess prospective operations. The latter proposed an evaluation of \gls*{conops} built upon two criteria. First, the \emph{scenario acceptability} which is referred to as \textit{the ability to effectively, efficiently, and reliably conduct operations with accurate exchange of all pertinent information and without excessive workload or (in-sim) avoidable inefficiencies or delay} \cite{Beaton2019Assessing}. The same criteria apply for system engineering when hardware or software are evaluated in their operational context \cite{SilvaMartinez2021HumanInTheLoop}. Second, the \emph{simulation quality} stands for the extent to which the simulation enables meaningful evaluation of the test objectives. One important aspect to highlight in simulation quality is fidelity, meaning the degree to which the simulation \textit{"replicates a real-world, built-environment from the user's perspective by considering form, function and user-interaction"} as defined by \citet{SilvaMartinez2021HumanInTheLoop}. It is oftentimes categorised into low-, mid- and high-level, and as the real world is the reference for fidelity in this context, perceived realism serves as a fundamental criterion behind this categorisation. Based on the literature, we propose that the fidelity of three \gls*{conops} key aspects is of critical importance to simulate these scenarios:

\begin{itemize}
    \item \textbf{Environmental Fidelity}, referring to how closely an analogue environment resembles the one it seeks to emulate. Specific locations on Earth can serve as high-fidelity analogues for extraterrestrial purposes \cite{Beaton2019Assessing}. However, such terrestrial analogues may also lack key lunar features (e.g., pitch black shadows, poor landscape cues \cite{Connors1994InterviewsDesign, Eppler1991LightingOperations}) and typically include elements not present in the environment being emulated, such as vegetation.
    \item \textbf{Mockup Fidelity}, relating to the degree of authenticity of the equipment mockups used in the simulation. According to \citet{SilvaMartinez2021HumanInTheLoop} \emph{low-fidelity} mockups should be used to assess anthropometric accommodation, placement and orientation of components, or even habitable volume. \emph{Medium-fidelity} mockups have proven useful to assess physical use and interaction with human interfaces (e.g., visibility, legibility, physical aspects of workload, usability, etc.). Finally, \emph{high-fidelity} mockups are to be used for test and training purposes. \citet{SilvaMartinez2021HumanInTheLoop} also argue that mockup fidelity should increase as the system design evolves.
    \item \textbf{Operations Fidelity}, referring to the fidelity of the performed tasks and underlying interactions. Whilst closely related to hardware fidelity, one could argue that performing analogue field study activities without considering the suit constraints, for example, can in many cases still yield sufficient insights. Such an approach can allow the review of relevant aspects of the operation without overburdening the study with suit components that may still be at an early stage of their development. \cite{Abercromby2013EvaluationExploration}. Furthermore, leaving out such elements does not prevent field-testing teams from assessing relevant risks during field crossing tasks \cite{Beaton2019UsingExploration}, which will be critical for future lunar surface scenarios \cite{Coan2020Conops}.
\end{itemize}

However, relying on analogue field studies and facilities has proven to induce recurrent cost overruns and planning delays due to substantial requirements in terms of human and logistic resources \cite{casini2020}.

Concerning the metrics for \gls*{conops} evaluation, subjective quantitative metrics are admittedly sufficient to assess certain aspects of \gls*{conops} but not robust enough to facilitate holistic assessment and revision of simulated procedures. It has therefore been argued they ought to be complemented with detailed qualitative feedback on the specific issues encountered \cite{Abercromby2013EvaluationExploration}. In other words, a phenomenological assessment in the form of subjective reflections is oftentimes indispensable to fully evaluate and improve \gls*{conops}. For example, feedback from the Apollo astronauts collected by \citet{Connors1994InterviewsDesign} spotlights that ergonomics of the spacesuit, especially the gloves dexterity, were in dire need of improvement. Objective data is also valuable to evaluate prospective tasks. For example, physiological data provides insights into ergonomics of spacecraft assembly tasks \cite{Osterlund2012VirtualTraining}, or retrospectively into the challenges of Apollo lunar surface activities \cite{Miller2017OperationalActivity}.

\subsection{Using Virtual Reality in Place of Analogues}

In contrast to analogue field and facility studies, immersive \gls*{vr} technologies appear to offer a new cost-effective and rapid prototyping medium to evaluate scenarios and the associated operational concerns \cite{Nilsson2023ShapeHumanity}. Indeed, while terrestrial analogues fail at simulating some of the critical extra-terrestrial features, such as the absence of atmospheric light scattering on the Moon \cite{Connors1994InterviewsDesign}, \gls*{vr} systems can easily recreate such conditions. Moreover, \gls*{vr} can, for example, visually and interactively simulate reduced gravity conditions. This per se does not automatically mean that \gls*{ive}s offer better environmental fidelity than analogues. Rather, they appear well predisposed for enabling evaluation of certain \gls*{conops} aspects. With regards to space applications, \gls*{vr} has already seen use as an operational planning tool \cite{Grubb2020UsingPole} or to assess the feasibility and difficulty of assembly and maintenance tasks \cite{Osterlund2012VirtualTraining}. For the maritime industry, \citet{Aylward2021Operations} successfully demonstrated the use of immersive environments to refine \gls*{conops}. Their methodology consisted in recreating extensively documented past accident scenarios in \gls*{vr}. They collected feedback from experts who completed the simulated scenario, which in turn provided the grounds for evaluation of potential hazards pertinent to a novel ship's bridge design. A potential drawback of this approach lies in the lack of interactivity between the experts and the environment. Based on their findings, the authors stated that they cannot rigorously verify that \gls*{vr} systems can evaluate \gls*{conops} and thus provide meaningful insights for this purpose. Consequently, even though \gls*{vr} shows great potential to help facilitate evaluations and assessments of \gls*{conops}, currently such an approach seems lacking empirical justification. 

Furthermore, \gls*{vr} is often criticised for its inherent lack of haptic feedback, which has been suggested to potentially undermine the validity of relevant design evaluations \cite{Nilsson2023ShapeHumanity}. Haptic interfaces therefore stand as potentially vital complements to interactive audio-visual simulations. In essence, they induce two types of haptic sensations; either tactile (i.e., of shape and texture through receptors in the skin) \cite{Lederman:2009:HapticsTutorial} or kinesthetic (i.e., of forces and body movements through muscular receptors, tendons and joints) \cite{Jones:2000:KinestheticSensing}. Classically, solutions to convey haptics in \gls*{vr} can be classified along a continuum spanning from active to passive haptics \cite{Zenner:2022:AdvancingProxyBasedHapticsInVR, Zenner:2017:Shifty}. Active haptics leverage actuators to actively exert forces on the user's body to simulate physical properties of objects inside the \gls*{ive}~\cite{Srinivasan:1997:Haptics}. This approach, however, suffers from being complex and costly, usually only works within constrained workspaces, and mostly fails to convey multimodal stimulations~\cite{Wang:2020:MultimodalVRHapticsSurvey}. Alternatively, passive haptics refrain from actuation but instead leverage the highly realistic and rich tactile and kinesthetic impressions provided by physical props in combination with the visual input of a VR system~\cite{Hinckley:1994:Props, Insko:2001:PHF, Nilsson:2021:ProppingUpVR}. According to the Haptic Fidelity Framework by \citet{Muender:2022:HapticFidelityFramework}, passive haptics can be considered a solution that is "low" in terms of "Versatility" but "high" in terms of "Haptic Fidelity". Its central advantage is that convincing multimodal feedback can be achieved at low costs using only low-fidelity and low-complexity mockups. Its main drawback, in turn, lies in the fact that props show poor flexibility to adapt to different situations -- a challenge that the community addressed by proposing mixed haptic interfaces~\cite{Zenner:2017:Shifty, Zenner:2019:Dragon, Mercado:2021:EncounteredTypeHapticsReview} and virtual illusion techniques like pseudo-haptics~\cite{Lecuyer2009SimulatingFeedback, Samad:2019:PseudoHapticWeight} or hand redirection~\cite{Azmandian:2016:Retargeting, Kohli:2013:Distortion, Zenner:2021:DPHFHR}. As such, passive haptics lends itself to use cases like specialist training, and thus appears highly suitable for VR-based ConOps assessment.

\subsection{Validity of Using Virtual Reality}

At least in theory, \gls*{vr} simulations thus appear to have the potential to substitute for physical simulations in certain situations. The study of this immersive medium has been a great endeavour over the past decades, resulting in the emergence of several models to characterise user experiences. Notably, improving the fidelity of VR-based simulations requires considering two aspects inherent to this medium: \textit{presence} and sense of \textit{embodiment}.

Presence in the context of \glspl*{ive} has originally been defined as the feeling of \textit{being there} in the simulated environment \cite{Slater1997AEnvironments, Witmer1998Presence}. It has also been described as the psychological response to experiencing immersion, meaning "\textit{the extent to which a display system can deliver an inclusive, extensive, surrounding, and vivid illusion of virtual environment to a participant}" \cite{Slater1999Response}. Nowadays, three presence dimensions are commonly considered \cite{Lee2004Presence}: physical, social and self-presence. Our focus here is mainly on physical presence or the feeling of experiencing the virtual world as the real one, also referred to as \textit{Place Illusion} by \citet{Slater2009PlaceIllusion}. The Igroup Presence Questionnaire (IPQ) introduced by \citet{Schubert2003IPQ} proposes to assess a sense of physical presence by means of the following sub-dimensions:
\begin{itemize}
    \item \textbf{Spatial Presence:} The sense of feeling physically present in the \gls*{ive}.
    \item \textbf{Sense of Realness:} The subjective experience of perceived realism in the \gls*{ive}, regarding all aspects it may encompass.
    \item \textbf{Involvement:} The high psychological state of attention paid to the \gls*{ive} \cite{Witmer1998Presence}.
\end{itemize}
The importance of presence for fidelity was already established by \citet{Slater2009PlaceIllusion}, who argued that feeling present while experiencing a plausible scenario, with respect to the proposed environment, results in a tendency of the user to react to virtual events as if they were real. Therefore, presence is praised for improving efficiency of training applications \cite{Piccione2019VirtualAgency}, which could also benefit \gls*{conops} assessment. Generally speaking, realism can contribute to the fidelity of simulations, and we argue that for VR-based simulations it must be considered through the prism of presence.

While the sense of physical presence relates to the environment, the sense of embodiment refers to the feeling that emerges when properties of the virtual body are processed as if they were properties of the user's biological body \cite{Kilteni2012Embodiment}. Three dimensions contribute to eliciting and maintaining a sense of embodiment toward a user's virtual representation:
\begin{itemize}
    \item \textbf{Self-Location:} The spatial experience of being inside the virtual body (not inside the virtual world) \cite{Debarda2015Perspective}.
    \item \textbf{Agency:} The sense of having full motor control over the virtual body and encompassing the intention of actions \cite{Blanke2009Bodyillusions, Jeunet2018MeasureAgency}.
    \item \textbf{Body Ownership:} The self-attribution that one makes of the virtual body \cite{Slater2009InducingOwnership, Kokkinara2019VBO}.
\end{itemize}
The contribution of the sense of embodiment to the fidelity of \gls*{vr} simulations stems from the fact that experiencing a strong embodiment reduces the gap between the user's virtual representation and its real-world counterpart. \citet{Gorisse2019Doppelganger}, for instance, witnessed that increasing ownership by using scanned avatars of the participants may incline them to protect more their bodily integrity against a threat.

Despite the senses of presence and embodiment being key drivers of simulation fidelity, their role in operational scenarios assessment remains underexplored. As mentioned before, traditional \gls*{vr} inherently lacks fidelity with regards to critical aspects of \gls*{conops} assessment, such as the ergonomics of lifting and manipulating physical objects. It would then seem reasonable to assume that this gap can be addressed by introducing haptic interfaces, as suggested by \citet{WeeHaptic2021}. Haptics are known to influence both presence and embodiment, which in turn may potentially impact some aspects of the simulation’s fidelity. Indeed studies report that passive haptics increase perceived presence \cite{Insko:2001:PHF, VicianaAbad2010HapticInteraction}, such as by increasing the perceived realism of objects that are not being interacted with \cite{Hoffman:1998:ProxyPlate}, and by increasing embodiment \cite{Medeiros2023BenefitsPassiveHaptics}. On the other hand, if the multimodal feedback afforded by the haptic proxy is not plausible in the virtual environment, this may impair presence \cite{Lindeman:1999:Dissertation}. Similarly, anthropomorphic similarities between actual and virtual representations may increase body ownership \cite{Argelaguet:2016:EffectsHandRealismOwnership}. In terms of embodiment, visuo-tactile correlation is a critical contributor to experiencing ownership \cite{Botvinick:1998:RubberHandIllusion, IJsselstejin2006RHI, Slater2009InducingOwnership}. Therefore, it can be reasonably theorised that discrepancies between worn passive haptic gears and user's virtual representation may also impair embodiment.

The concepts of presence and embodiment thus constitute critical elements influencing simulation fidelity. The research community would consequently benefit from investigating the potential frictions and synergies between these concepts and haptics for operational scenarios evaluation.

In summary, this literature review explores current methods used to assess \gls*{conops} in the context of human space exploration. Traditionally, this involves expensive and logistically intensive analogue field studies that aim to replicate key aspects of space operations with a high fidelity, allowing for valid evaluation of potential scenarios.  Due to their capacity to simulate key aspects of space environments at a comparatively low cost, immersive \gls*{vr} simulations are emerging as a potentially viable alternative. Whilst VR is generally limited to conveying audiovisual experiences, existing literature suggests this might be compensated for using haptic interfaces capable of delivering tactile and kinesthetic sensations. 
A responsible use of such multimodal \gls*{vr} interfaces in \gls*{conops} assessments will require careful consideration of the psychological states they induce in participants, namely the senses of presence and embodiment, both of which have been found to impact the fidelity of \gls*{vr} simulations in significant ways. 
The following section outlines our strategy to enhance the fidelity of a \gls*{vr} simulation by employing passive haptic proxies. With that we aim to shed further light on their viability and prospective role in upcoming operational assessments.

\section{Experiment}\label{section:user study}

\subsection{The Apollo 12 Case Study}


In order to assess and validate a VR simulation of a lunar ConOps, it is crucial that we compare it to a real-world counterpart. We thus turned to the Apollo program. As an example, the historical Apollo 12 mission entailed a two-man surface crew, composed of Charles Conrad Jr. and Alan L. Bean, deploying science equipment and collecting sampless. We turned attention to the \gls*{eva} focusing on deploying the first iteration of the \gls*{alsep} designed by the company Bendix Aerospace. \gls*{alsep} consisted of 2 packages hosting 6 deployable surface experiments, a central station to administrate the overall working and data transfer, along with a shared power supply in the form of the \gls*{rtg}. Its goal was to monitor many characteristics of the lunar environment. Its deployment followed a strict written procedure for which the crew was intensively trained on Earth. Key steps of this procedure were focusing on retrieving packages from the lander, attaching handling tools to them for easier manipulation, fuelling the \gls*{rtg}, and assembling both packages into a barbell to facilitate their transport by one crew member to the final drop-off point. Upon completing its transportation, the crew had to set up scientific equipment at the selected deployment site. In spite of having taken place more than half a century ago, the \gls*{alsep} deployment remains one of the most extensively documented lunar surface operations, with substantial amounts of authentic astronaut feedback being available to the public (see \autoref{tab: Apollo feedback}). Moreover, the diverse range of actions involved, such as identification, grabbing and assembling pieces of equipment, along with its representative nature with respect to future missions, makes the \gls*{alsep} deployment procedure into a suitable baseline for our study. We therefore recreated part of the extensively documented Apollo 12 mission in VR with the aim of comparing feedback from Apollo astronauts with that of experts in human spaceflight experiencing the VR simulation, similarly to the methodology of \citet{Aylward2021Operations}. The actual \gls*{eva} lasted for 4 hours, but immersing our participants in \gls*{vr} for such a long period of time was not an option, thus the simulated analogue mission was reduced to the following outlined procedure:
\begin{enumerate}
    \item Retrieving package \#1 from the lander’s cargo bay to drop it at a designated location on the surface.
    \item Retrieving package \#2 from the lander’s cargo bay to drop it at a designated location on the surface.
    \item Retrieving a specified handling tool from package \#1 and attach it to package \#1 to facilitate its future manipulation.
    \item Retrieving a specified handling tool from package \#1 and attach it to package \#2 to facilitate its future manipulation.
    \item Retrieving both parts of the carry mast from package \#1, assembling them as a whole bar.
    \item Assembling both packages \#1 \& \#2 as a barbell by attaching one package at each tip of the carry bar.
    \item Transporting the assembled barbell to the designated drop-off point.
\end{enumerate}

\begin{figure*}[!ht]
    \centering
    \includegraphics[width=\textwidth]{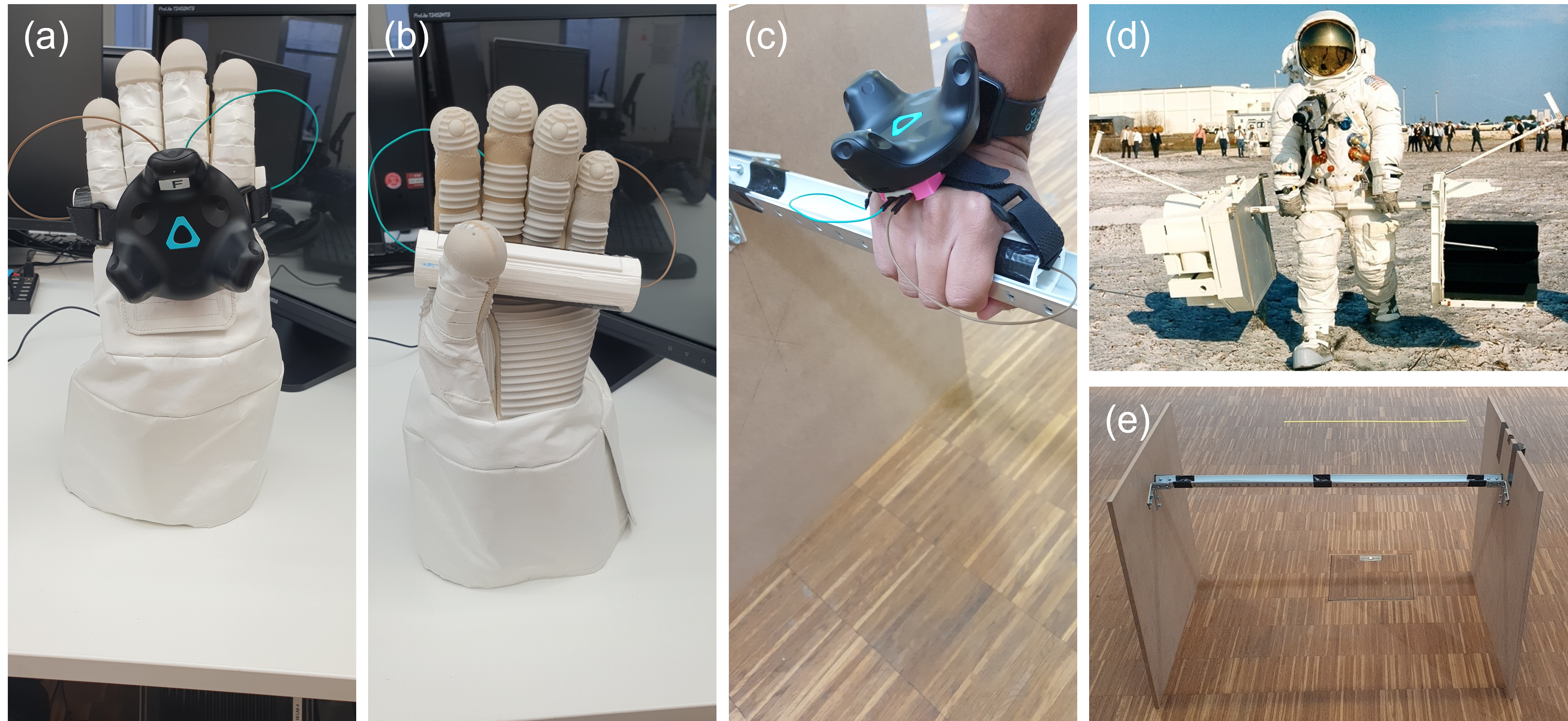}
     \caption{(a) Back and (b) front views of the gloves with our controller. (c) The controller nicely fits inside the mockup's carry mast when grasped, completing its shape to a tube. (d) High-fidelity mockup of the carry barbell as used in real-world analogue field studies (credit: NASA). (e) Low-fidelity mockup of the barbell used in the VR field study.}
     \label{fig:collage}
     \Description{This figure is a collage of the haptic props used for the study: an astronaut glove equipped with the hand controller, from front and back views; the hand controller fitting in the Barbell physical mockup; high and low fidelity versions of barbell mockups.}
\end{figure*}

\subsection{Virtual Simulation}

A virtual lunar environment was created using Unreal Engine 4. To maximise authenticity, we used topographic scans of the Lunar surface, taken by the \gls*{lroc} at the exact location of the Apollo 12 mission landing site. Ephemerides from \gls*{nasa} \gls*{jpl} allowed us to recreate the same lighting conditions as they were at the mid-time of the Apollo 12 \gls*{eva}. A high-fidelity 3D model of the \gls*{lem} was placed at the border of the crater where the crew landed and oriented into the scene based on accurate Apollo 12 data. 

Participants were embodied in a virtual representation of the Apollo suit’s gloves as well as in a virtual helmet reproducing the restricted field of view of an actual suit. Previous works suggest partial body representations can elicit a strong sense of embodiment \cite{Argelaguet:2016:EffectsHandRealismOwnership, Dewez:2021:AvatarFriendly3DManipulationTechniques, Eubanks:2020:EffectsBodyTrackingFidelityEmbodiment}. A helmet flashlight could be toggled by the participants at their convenience, by triggering an input with their right hand close to their virtual astronaut helmet. For the scenario to be performed by one single user, as a first proof of concept, we implemented part of the initial single-crew deployment procedure written by Bendix Aerospace \cite{Marrus1968ALSEP}. Despite this procedure being actually performed by a two-man crew, we argue that simulating it as a single-crew deployment for the purpose of early \gls*{conops} assessment remains valid. Especially, the unitary tasks introduced above were only performed by one crew member at a time during the actual scenario. Visual indicators of the elements were used to guide the participants through the deployment procedure.

To populate the \gls*{ive}, both \gls*{alsep} science packages and their complete set of tools were accurately recreated as virtual models. Interactions between the different components happened automatically thanks to a snapping feature, once all elements were in the vicinity of the trigger zone. Also, the physics of the virtual environment was tweaked to emulate the Moon's gravity. All interactable assets were then impacted by the game engine physics, except the barbell representation when tracked by the physical prop. Finally, the package drop-off location was marked by a flag.

\subsection{Apparatus}

The audiovisual apparatus consisted of a HTC Vive Pro to display the virtual environment at a refresh rate of 90 Hz with a resolution of 2880 × 1600 pixels (1440 × 1600 pixels per eye) and providing the user with a diagonal \gls*{fov} of 110 degrees. \gls*{vr} hardware was powered by a desktop computer equipped with a Nvidia RTX 2080 Ti GPU, an Intel i9-12900K CPU and 32Gb of RAM.

A medium-fidelity reproduction of actual \gls*{eva} gloves was employed as a wearable passive haptic interface to recreate the movement restrictions that characterised real gloves \cite{Schmitt_2009}. In spite of lacking full pressurisation of the authentic astronaut gloves, their rubber material was deemed stiff enough to give a sense of the movement restrictions. As spacesuit design continues to evolve, several ongoing initiatives are in the process of exploring prospective solutions tailored to the lunar context. Notably, the AxEMU spacesuit developed by Axiom Space promises a “superior fit for astronauts while increasing their comfort and ability to perform tasks” \cite{axiom2023}. It is consequently likely that future astronaut gloves might exhibit haptic properties differing from our prototype in ways that are difficult to predict at this point. While any attempts at broad generalisations based on our glove-related findings should thus be approached cautiously, it is worth noting that our setup is representative and relevant to current operations from a usability standpoint.

Additionally, we wanted to investigate a second type of passive haptic interface for \gls*{conops} assessment, i.e., a handheld physical proxy. To that end, a low-fidelity mockup of the assembled barbell used during the Apollo 12 mission was designed (see \autoref{fig:collage}(d)(e)). The actual barbell would have a mass of around $163~kg$, resulting in a weight roughly equivalent to $27~kg$ on the Moon. Some mass was therefore added to our mockup to weigh it up to $27~kg$. The position of the virtual barbell was kept in sync with its physical counterpart thanks to a regular Vive Controller attached to the latter (see \autoref{fig:teaser}).

\begin{figure}[t]
    \centering
    \includegraphics[width=\linewidth]{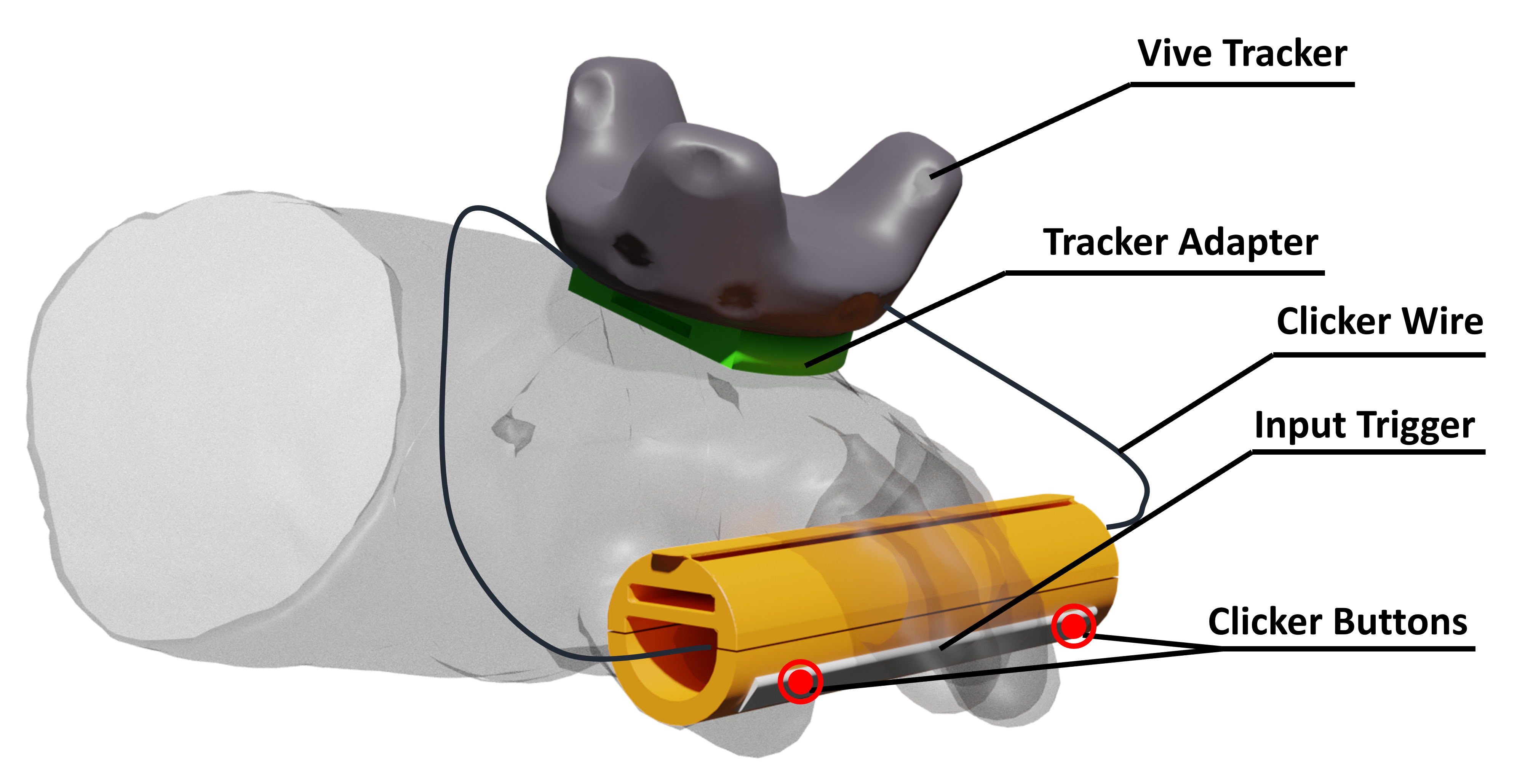}
     \caption{Outline of the hand controller features}
     \label{fig:hand controller}
     \Description{Shows the features of the hand controllers, mainly the input trigger and the integration of the vive tracker.}
\end{figure}

To allow for seamless grabbing of virtual objects in the simulated space, innovative hand controllers were designed to enable manipulation of the physical mockup in an unobtrusive manner. To that end, our controllers took on the form of a tubular shape representative of most of the manipulated Apollo tools' handles. The controllers were strapped to the user’s hand, as depicted in \autoref{fig:collage}. The design is outlined on  \autoref{fig:hand controller} and features two clickable buttons to trigger a binary grabbing action in the \gls*{ive}. The virtual hand would close on press of a clicker and opening at release. Both clickers are embedded behind a bar-like button and each of them could trigger an input through being wired to the POGO-pins of the HTC Vive Tracker. The latter is attached to the user's wrist or glove and is thus in charge of transmitting grabbing inputs to the simulation and tracking the hand. Finally, due to their shape, our bar-controllers nicely fit into the gutter-like shape of the mockup’s carry mast, completing it into an actual tube similar to the actual carry bar.

\subsection{Experimental Design}

We used two variables for this experiment following a mixed (between-within-subjects) design:
\begin{itemize}
    \item \textbf{Gloves (within-subjects variable):} The aforementioned reproduction of actual astronaut gloves were either worn (Gloves) or not (NoGloves) during the session to allow participants to compare experiences. These conditions were counterbalanced to cope with habituation effects.
    \item \textbf{Barbell Mockup (between-subjects variable):} The low-fidelity barbell mockup based on the Apollo 12 equipment was used by half of the participants in the passive haptic barbell proxy condition (Mckp), while the other half had no haptic interface to perform the carrying task (NoMckp).
\end{itemize}

\subsection{Participants}

\begin{table*}[t]
    \centering
    \resizebox{0.8\linewidth}{!}{%
    \begin{tabular}{llll}
    \hline
        \textbf{Participant category} & \textbf{Gender} & \textbf{Job title} & \textbf{Domain of expertise} \\ \hline 
        Astronaut1* & F & Astronaut &  Human Spaceflight, EVA \\
        Astronaut2* & M & Astronaut &  Human Spaceflight, EVA \\
        AsCan* & M & Astronaut Candidate & Human Spaceflight, EVA  \\ \hline
        Engineer1 & M & Extended Reality Engineer & Interactive Media, XR Training  \\
        Engineer2 & M & Extended Reality Engineer & Haptics, Robotics, XR Training  \\
        Engineer3 & M & Extended Reality Engineer & Software Engineering, XR Training \\
        Engineer4 & F & Ops Engineer & Ergonomics, Analogue Field Ops \\
        Engineer5 & M & Aerospace Engineer & In Situ Resource Utilisation \\
        Engineer6 & M & Extended Reality Engineer & Haptics, XR Training \\
        Engineer7 & M & Aerospace Engineer & In-Space manufacturing, Advanced Manufacturing\\
        Engineer8 & M & Crew Support Engineer and EUROCOM & Physics \\
        Engineer9 & M & Science and Ops Engineer & Facility Activities Management \\ \hline
        Instructor1 & M & Astronaut Instructor & EVA training  \\
        Instructor2 & F & ISS Astronaut Instructor and EUROCOM & Columbus and Payload Instructor, Astrophysics \\
        Instructor3 & M & ISS Astronaut Instructor and EUROCOM & ISS Payload Instructor  \\ 
        Instructor4 & M & Astronaut Trainer and Ops Engineer &  Procedure Management \\ \hline
        Manager1 & M & Management Support & Human Spaceflight, Aerospace Engineering  \\
        Manager2 & M & Project Manager & Post-ISS Ops, Ground Segment and Project Engineer  \\ \hline
        Scientist1 & F & Research Fellow  & Space Medicine  \\
        Scientist2 & F & Research Fellow & Machine Learning, Planetary Science  \\
        Scientist3 & M & Research Fellow & Artificial Intelligence  \\
        Scientist4 & M & Research Fellow & Space Radiation Shielding  \\
        Scientist5 & M & Research Fellow & Scientific Research and Development  \\\hline
        Student1 & M & Extended Reality Trainee & Augmented Reality, Software Engineering  \\
        Student2 & F & Extended Reality Trainee & Cognitive Sciences  \\
        Student3 & M & Aerospace Engineering Trainee & Spacecraft Systems  \\
        Student4 & F & Visiting Trainee & Data Science  \\ \hline
    \end{tabular}%
    }
    \caption{Summary of the participants' profiles. The indicated job titles and domains of expertise were self-reported (*withdrawn from the quantitative analysis)}.
    \label{tab: participants}
    \Description{A summary of the participants profiles, comprising 3 astronauts, 9 engineers, 4 instructors, 2 managers, 5 scientists and 4 students with an expertise in human spaceflight.}
\end{table*}

Given that our investigation is quite domain-specific, we handpicked and extended personalised invitations to relevant experts in the field of human spaceflight. Overall, 24 participants (6 females and 18 males) aged from 22 to 60\footnote{$M = 34.12$, $SD = 10.99$} took part in the quantitative study. Table \ref{tab: participants} summarises our panel of participants. 2 career astronauts who are still active were included in our study. Put together, they have logged around 547 days in space and have also totalled 14h of \gls*{eva} operations outside the \gls*{iss}. One astronaut candidate (AsCan) who has already started his basic training also took part in the study. Unfortunately, Astronaut 1 was not able to complete the whole planned study due to schedule constraints, and along with Astronaut 2 and AsCan they performed the study with an earlier iteration of the controllers' prototype, which used a different input schema that did not meaningfully impact the experience, but might have impacted some of the quantitative measurements. We thus removed their answers from the quantitative dataset and therefore from the participants descriptive statistics, while still considering their feedback in the qualitative analysis. Engineers were recruited from diverse fields of expertise, but all related to space mission support and training activities. The participating instructors were all deeply experienced with delivering astronaut training, some of them being also involved in astronaut mission activities through the \gls*{eurocom} which is the centre responsible for communication with the \gls*{iss}. Most of the scientists were involved in field studies activities for future lunar and martian exploration missions, while others worked on topics related to Moon exploration. The managers were involved in human spaceflight programmes from a support and administrative perspective. Finally, we extended invitations to students having relevant experience in the field of space exploration.

\subsection{Procedure and Measurements}

\subsubsection{Pre-experiment procedure}
Participants were first invited to fill in a demographic questionnaire and were asked to self-report about their gaming habits and \gls*{vr} experience to ensure there were no subsequent differences within our groups of participants. Indeed, our groups shared similar mean ages, levels of gaming habits and \gls*{vr} experience (see \autoref{tab: demo descriptive statistics}). No significant difference were thus observed between the groups. We can therefore guarantee their homogeneity between groups on this aspect that could affect participants' experience of the simulation. They were then shown a short briefing video providing an overview of the simulated mission. Finally, depending on their group, they were introduced to the physical mockup and were given relevant safety recommendations.

\subsubsection{Experiment procedure and tasks}
Participants were geared up and immersed in the virtual lunar environment. Prior to doing any activity in the virtual space, a calibration phase of the virtual gloves needed to be undertaken. It consisted of spatially registering the virtual gloves with the real ones by placing the real hands and fingers at the same location as their virtual counterparts. Once this phase was completed, the participants were guided through the \gls*{alsep} deployment process. Then, depending on their group, they were either asked to grab the fully virtual or physical mockup in order to transport it to the drop-off point about 91 m (300 ft) far from the lander. During this phase, participants were asked to stand while their virtual representation was "sliding" continuously and automatically in the direction of the user's gaze. Participants could at any time drop the barbell (virtual or physical) to stop moving and rest. 

Following a think-aloud approach \cite{Jaaskelainen2002ThinkAloud}, we asked participants to comment on any aspect of their experience they found noteworthy. In addition, we enquired about some of the mission design aspects identified from the Apollo 12 crew debriefing records \citep{NASA1968Debriefing}: the manipulation of experiment packages and associated tools, the transportation of the assembled barbell to the deployment area and thought-out improvements for the \gls*{conops}. The set of questions used as a common baseline among participants is reported in \autoref{Appendix: think aloud}. Particular attention was paid to formulating these questions in such a way that they addressed both the use of the VR interface and the mission objectives. Interviews were recorded and transcribed.

\subsubsection{Post-experiment procedure}
Upon completing a first \gls*{vr} session, participants were invited to fill in post-experiment questionnaires. The observed dependent variables are:
\begin{itemize}
    \item \textbf{Embodiment}, as a contributor to the perceived fidelity of the \gls*{vr} simulation, through the ownership and agency scales from the \gls*{veq} \cite{Roth2020VEQ}. The "change" item of the \gls*{veq} was not measured has no morphological modification of the virtual body was induced. Self-location was not considered as previous work give no clue it could be influenced by the props used.
    \item \textbf{Presence}, as a contributor to the perceived fidelity of the \gls*{vr} simulation, through the general presence, spatial presence and realness items from the \gls*{ipq} \cite{Schubert2003IPQ}. Involvement was not measured as its contribution to experiencing presence is debated \cite{Slater2003Note}.
    \item \textbf{System Usability}, to assess the global acceptance of our haptic configurations, through the \gls*{sus} \cite{Brooke1996SUS}.
    \item \textbf{Task load}, when manipulating objects and carrying the barbell, through the \gls*{nasa} \gls*{rtlx} \cite{Hart2006TLX}, in order to support comparison with the actual Apollo 12 crew feedback.
\end{itemize}
All metrics were assessed on 7-point Likert scales except for the system usability items that were on a 5-point one, accordingly to its guidelines. Participants were also able to freely express thanks to open-ended items in the questionnaire.

They were then invited to perform another \gls*{vr} session to experience the second glove condition and were finally asked to fill in another instance of the post-experiment questionnaires.

\subsubsection{Hypotheses}
In light of the literature review, we investigated four main hypotheses:

\begin{itemize}
    \item[\textbf{H1:}] Introducing passive haptics through wearing proxy gloves coherent with the virtual gloves increases embodiment through the sense of body ownership.
    \item[\textbf{H2:}] Introducing passive haptics (H2.1) through wearing proxy gloves coherent with the virtual gloves, and (H2.2) through using a physical mockup of the virtual barbell, increases the sense of presence.
    \item[\textbf{H3:}] Introducing passive haptics (H3.1) through wearing proxy gloves coherent with the virtual gloves, and (H3.2) through using a physical mockup of the virtual barbell, increases the global workload of the participants for the grabbing tasks.
    \item[\textbf{H4:}] Introducing passive haptics (H4.1) through wearing proxy gloves coherent with the virtual gloves,  and (H4.2) through using a physical mockup of the virtual barbell, increases the global workload of the participants for the field crossing task to the drop-off point while carrying the barbell.
\end{itemize}
\section{Results}\label{section: results}

\subsection{Quantitative Analysis}

\begin{table}[t]
	\centering
    \small
	{
		\begin{tabular}{lp{0.3in}p{0.3in}p{0.3in}p{0.3in}}
			\toprule
            \multicolumn{1}{c}{} & \multicolumn{2}{c}{No Mockup} & \multicolumn{2}{c}{Mockup} \\
			\cline{2-3}\cline{4-5}
            & $\Bar{x}$ & $\sigma$ & $\Bar{x}$ & $\sigma$\\
            \cmidrule[0.4pt]{1-5}
            Age                 & 32.4 & 12.1 & 35.8 & 10.0 \\
            Gaming frequency (0-5) & 1.5 & 1.68 & 1.75 & 1.66 \\
            VR experience & 4.25 & 1.86 & 3.83 & 1.99\\
			\bottomrule
		\end{tabular}
	}
    \caption{Demographic descriptive statistics}
    \label{tab: demo descriptive statistics}
    \Description{The demographic descriptive statistics presenting age, gaming frequency and virtual reality experience of the participants grouped by mockup or no mockup condition.}
\end{table}

\begin{figure}[ht]
     \centering
     \begin{subfigure}[b]{0.45\linewidth}
         \centering
         \includegraphics[width=\textwidth]{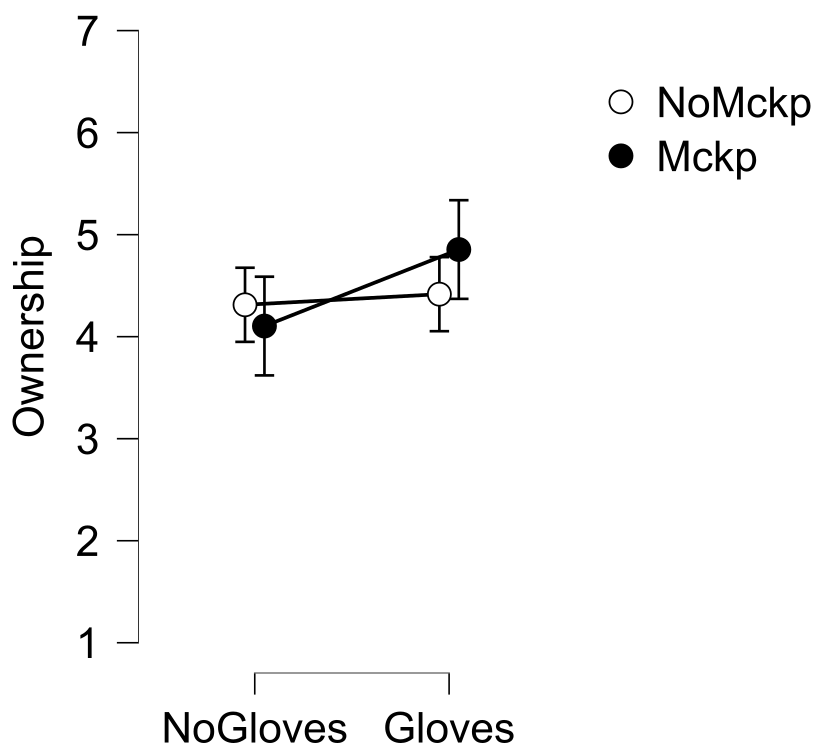}
         \caption{Ownership}
         \label{subfig: ownership plot}
     \end{subfigure}
     \hfill
     \begin{subfigure}[b]{0.45\linewidth}
         \centering
         \includegraphics[width=\textwidth]{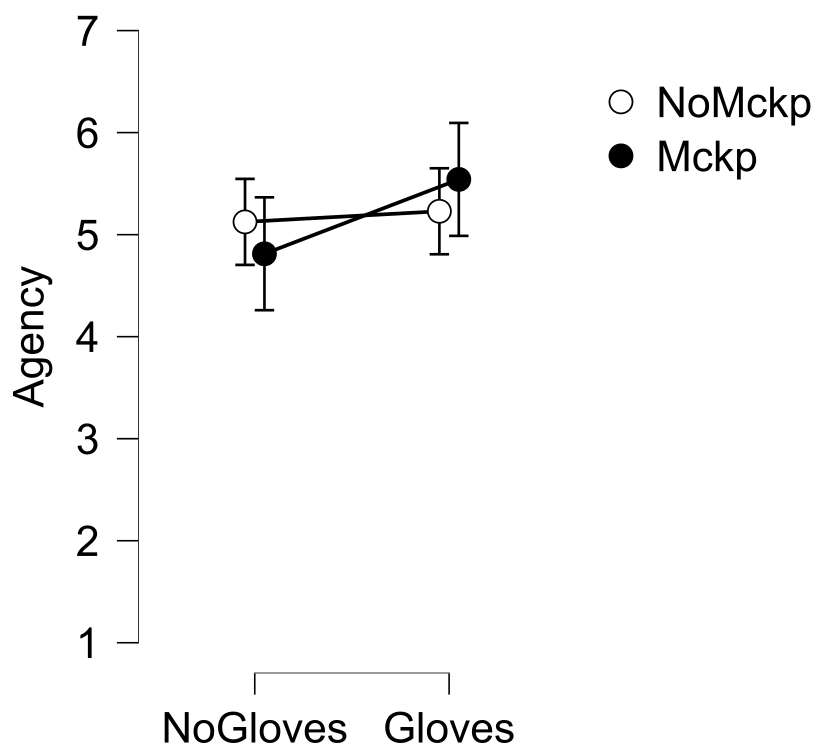}
         \caption{Agency}
         \label{subfig: agency plot}
     \end{subfigure}
     \begin{subfigure}[b]{0.45\linewidth}
         \centering
         \includegraphics[width=\textwidth]{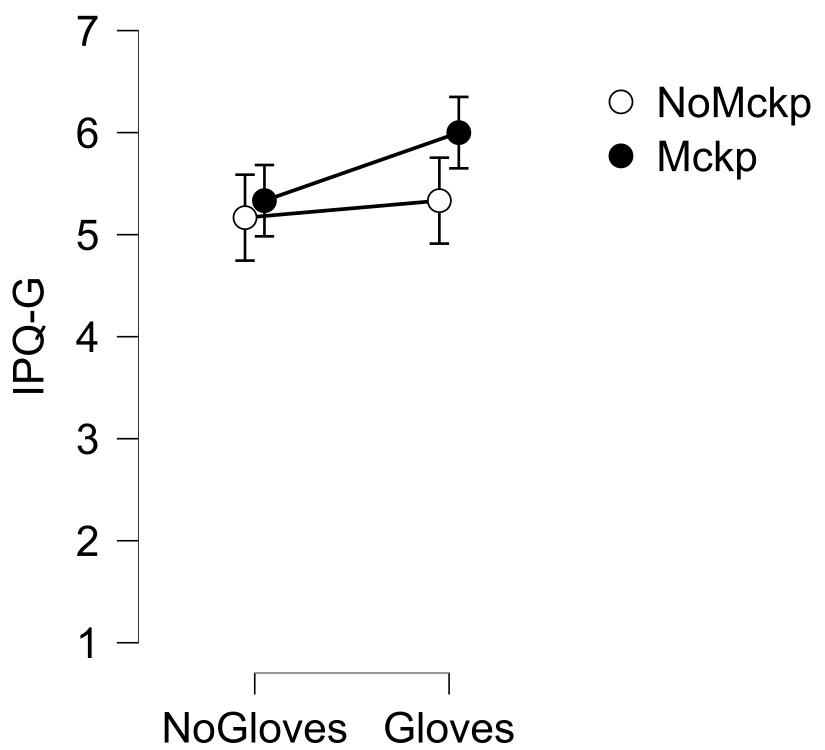}
         \caption{IPQ: general presence}
         \label{subfig: general presence plot}
     \end{subfigure}
     \hfill
     \begin{subfigure}[b]{0.45\linewidth}
         \centering
         \includegraphics[width=\textwidth]{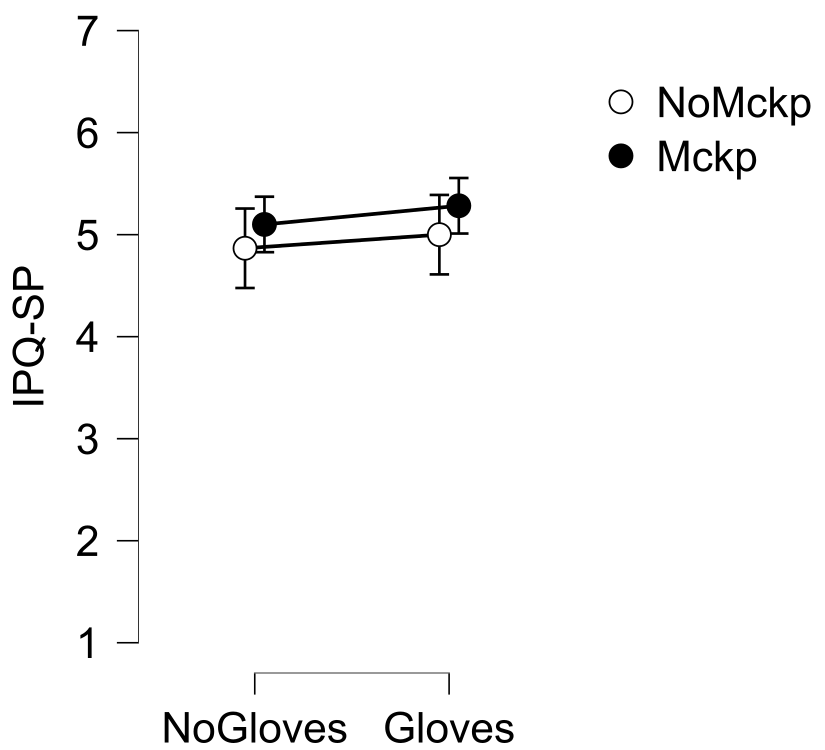}
         \caption{IPQ: spatial presence}
         \label{subfig: spatial presence plot}
     \end{subfigure}
     \begin{subfigure}[b]{0.45\linewidth}
         \centering
         \includegraphics[width=\textwidth]{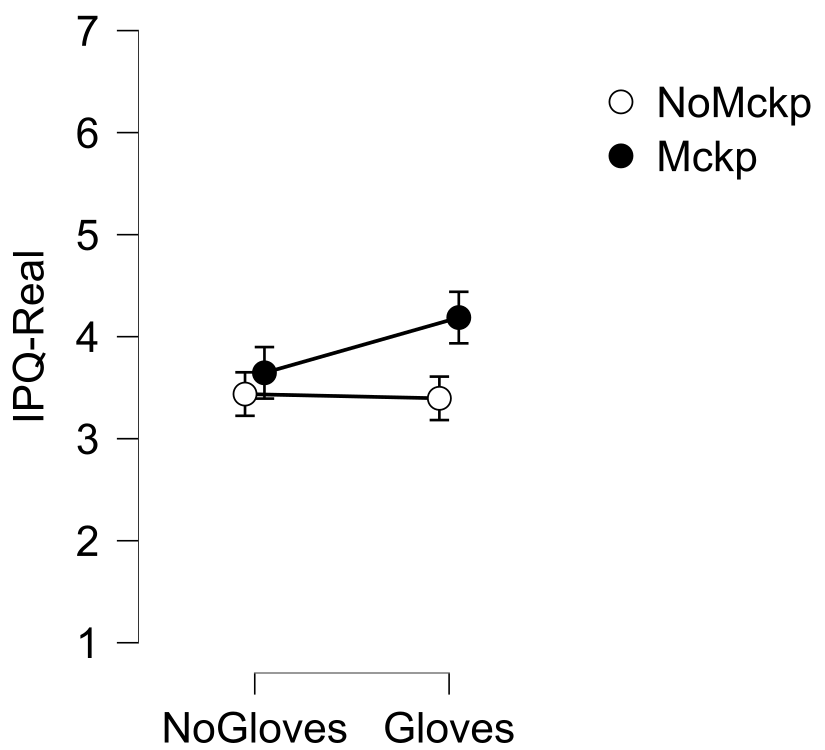}
         \caption{IPQ: realness}
         \label{subfig: Realness plot}
     \end{subfigure}
     \hfill
     \begin{subfigure}[b]{0.45\linewidth}
         \centering
         \includegraphics[width=\textwidth]{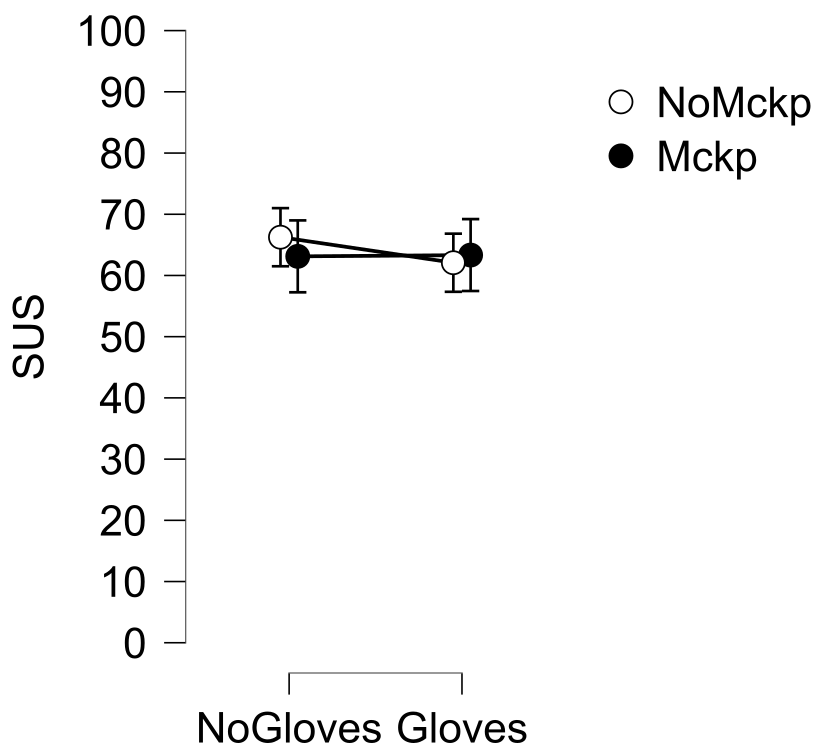}
         \caption{System usability}
         \label{subfig:SUS plot}
     \end{subfigure}
     \begin{subfigure}[b]{0.45\linewidth}
         \centering
         \includegraphics[width=\textwidth]{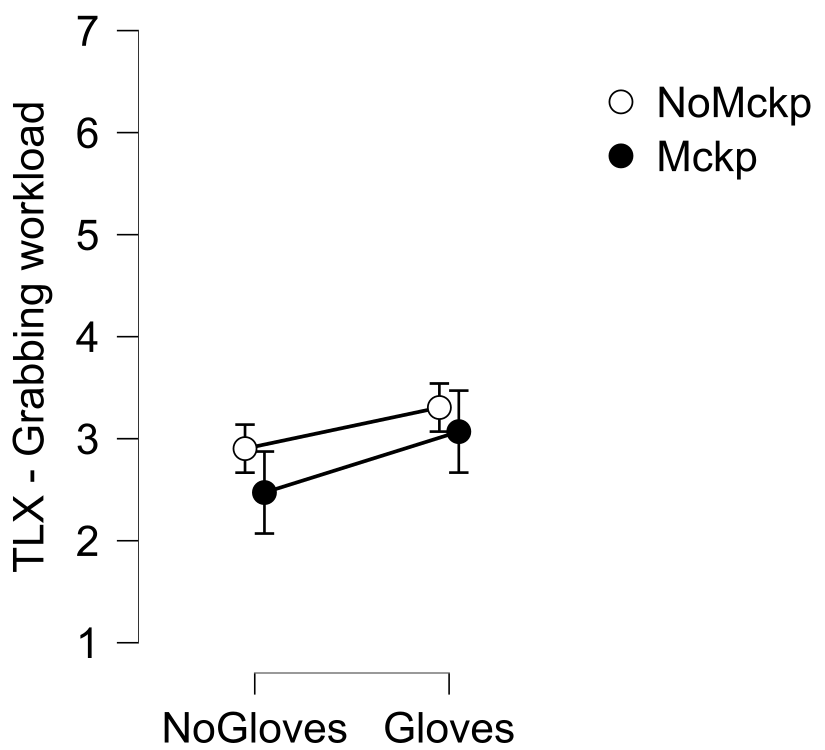}
         \caption{TLX: objects grabbing task}
         \label{subfig: tlx grabbing}
     \end{subfigure}
     \hfill
     \begin{subfigure}[b]{0.45\linewidth}
         \centering
         \includegraphics[width=\textwidth]{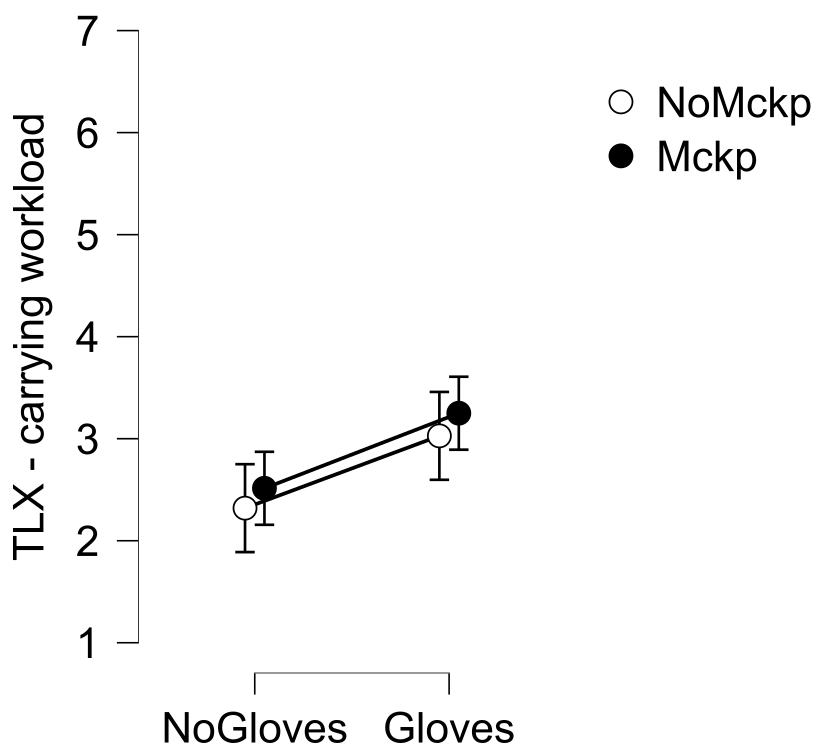}
         \caption{TLX: barbell carrying task}
         \label{subfig: tlx carrying}
     \end{subfigure}
        \caption{Descriptive plots}
        \label{fig:descriptive plots}
        \Description{The descriptive plots for the dependent variables.}
\end{figure}

\begin{table*}[!ht]
	\centering
    \small
	{
		\begin{tabular}{lp{0.3in}p{0.3in}p{0.3in}p{0.3in}p{0.3in}p{0.3in}p{0.3in}p{0.3in}}
			\multicolumn{1}{c}{} & \multicolumn{4}{c}{\textbf{No Mockup}} & \multicolumn{4}{c}{\textbf{Mockup}} \\
			\cline{2-5}\cline{6-9}
            \multicolumn{1}{c}{} & \multicolumn{2}{c}{\textbf{No Gloves}} & \multicolumn{2}{c}{\textbf{Gloves}} & \multicolumn{2}{c}{\textbf{No Gloves}} & \multicolumn{2}{c}{\textbf{Gloves}} \\
			\cline{2-5}\cline{6-9}
			 & $\Bar{x}$ & $\sigma$ & $\Bar{x}$ & $\sigma$ & $\Bar{x}$ & $\sigma$ & $\Bar{x}$ & $\sigma$  \\
            \cmidrule[0.4pt]{1-9}
		  \textbf{Ownership}         & 4.31  & .724 & 4.42  & .756 & 4.10  & 1.61 & 4.85  & 1.07 \\
            Agency           & 5.12  & .938 & 5.23  & .932 & 4.81  & 1.19 & 5.54  & .753 \\
            \cmidrule[0.4pt]{1-9}
            \textbf{General presence}      & 5.17  & 1.53 & 5.33  & 1.23 & 5.33  & 1.43 & 6.00  & .853 \\
            Spatial presence     & 4.87  & .924 & 5.00  & .840 & 5.10  & 1.11 & 5.28  & .916 \\
            \textbf{Realness}             & 3.44  & .806 & 3.40  & .661 & 3.65  & 1.07 & 4.19  & .820 \\
            \cmidrule[0.4pt]{1-9}
            System usability        & 66.2 & 11.5 & 62.1 & 13.5 & 63.1 & 19.7 & 63.3 & 20.2 \\
            \cmidrule[0.4pt]{1-9}
            \textbf{(Grabbing) TLX} & 2.90  & .553 & 3.30  & .531 & 2.47  & .855 & 3.07  & 1.06 \\
            (Grabbing) Mental demand    & 3.00  & 1.13 & 3.17  & 1.11 & 2.58  & 1.44 & 2.58  & 1.44 \\
            \textbf{(Grabbing) Physical demand}    & 1.67  & .888 & 3.50  & 1.68 & 1.83  & .577 & 3.75  & 1.54  \\
            \textbf{(Grabbing) Time demand}    & 3.42  & .793 & 3.50  & 1.24 & 2.50  & 1.17 & 2.58  & 1.50 \\
            (Grabbing) Performance    & 3.25  & 1.21 & 3.50  & 1.24  & 3.00  & 1.54 & 3.75  & 1.48 \\
            (Grabbing) Effort     & 3.67  & 1.07 & 3.92  & 1.24  & 2.75  & 1.54 & 3.42  & 1.78 \\
            (Grabbing) Frustration     & 2.41  & 1.16 & 2.25  & 1.05 & 2.17  & 1.59 & 2.33  & 1.67 \\
            \cmidrule[0.4pt]{1-9}
            \textbf{(Carrying) TLX} & 2.32  & .845 & 3.03  & .825  & 2.51  & .869 & 3.25  & 1.20 \\
            \textbf{(Carrying) Mental demand}    & 2.25  & .866 & 3.00  & 1.35 & 1.75  & .754 & 2.67  & 1.43  \\
            \textbf{(Carrying) Physical demand}    & 1.58  & .900 & 3.67  & 1.83 & 3.67  & 1.30 & 5.00  & 1.71 \\
            (Carrying) Time demand    & 3.50  & 1.45 & 3.75  & 1.36 & 3.25  & 1.86 & 2.83  & 1.70 \\
            (Carrying) Performance    & 2.50  & 1.45 & 2.50  & 1.38 & 2.17  & 1.11 & 2.92  & 1.56 \\
            \textbf{(Carrying) Effort}     & 2.33  & 1.15 & 3.33  & 1.37  & 2.50  & 1.31 & 3.92 & 1.73 \\
            (Carrying) Frustration     & 1.75  & 1.21 & 1.92  & .793 & 1.75  & .866 & 2.17  & 1.47 \\
			\bottomrule
		\end{tabular}
	}
    \caption{Descriptive statistics for the dependent variables.}
    \label{tab: DV descriptive statistics}
    \Description{The descriptive statistics for the observed dependent variables.}
\end{table*}

\begin{figure*}[!ht]
     \centering
     \begin{subfigure}[b]{0.42\linewidth}
         \centering
         \includegraphics[width=\textwidth]{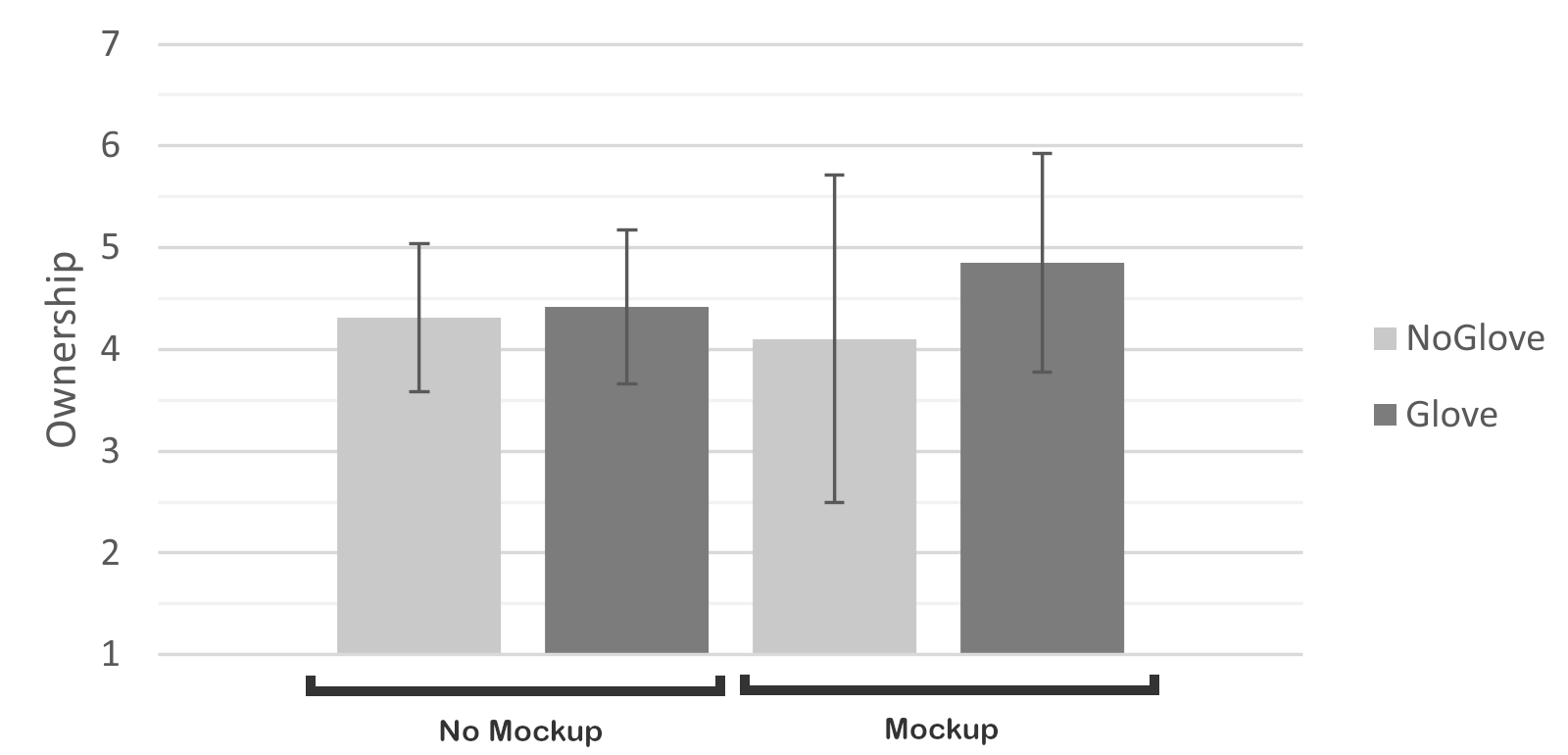}
         \caption{Ownership}
         \label{Barchart: ownership}
     \end{subfigure}
     \begin{subfigure}[b]{0.42\linewidth}
         \centering
         \includegraphics[width=\textwidth]{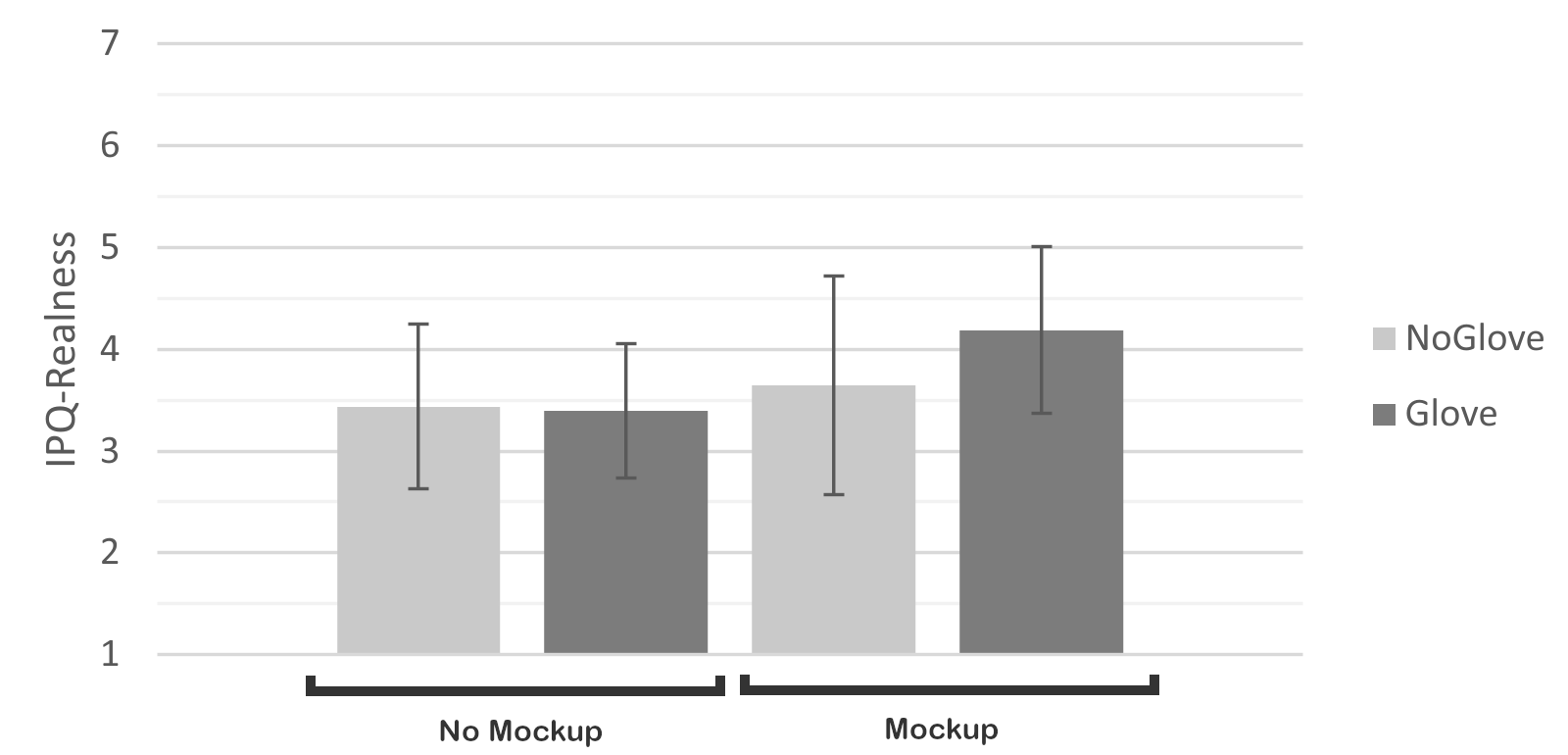}
         \caption{IPQ: realness}
         \label{Barchart: Realness}
     \end{subfigure}
     \begin{subfigure}[b]{0.42\linewidth}
         \centering
         \includegraphics[width=\textwidth]{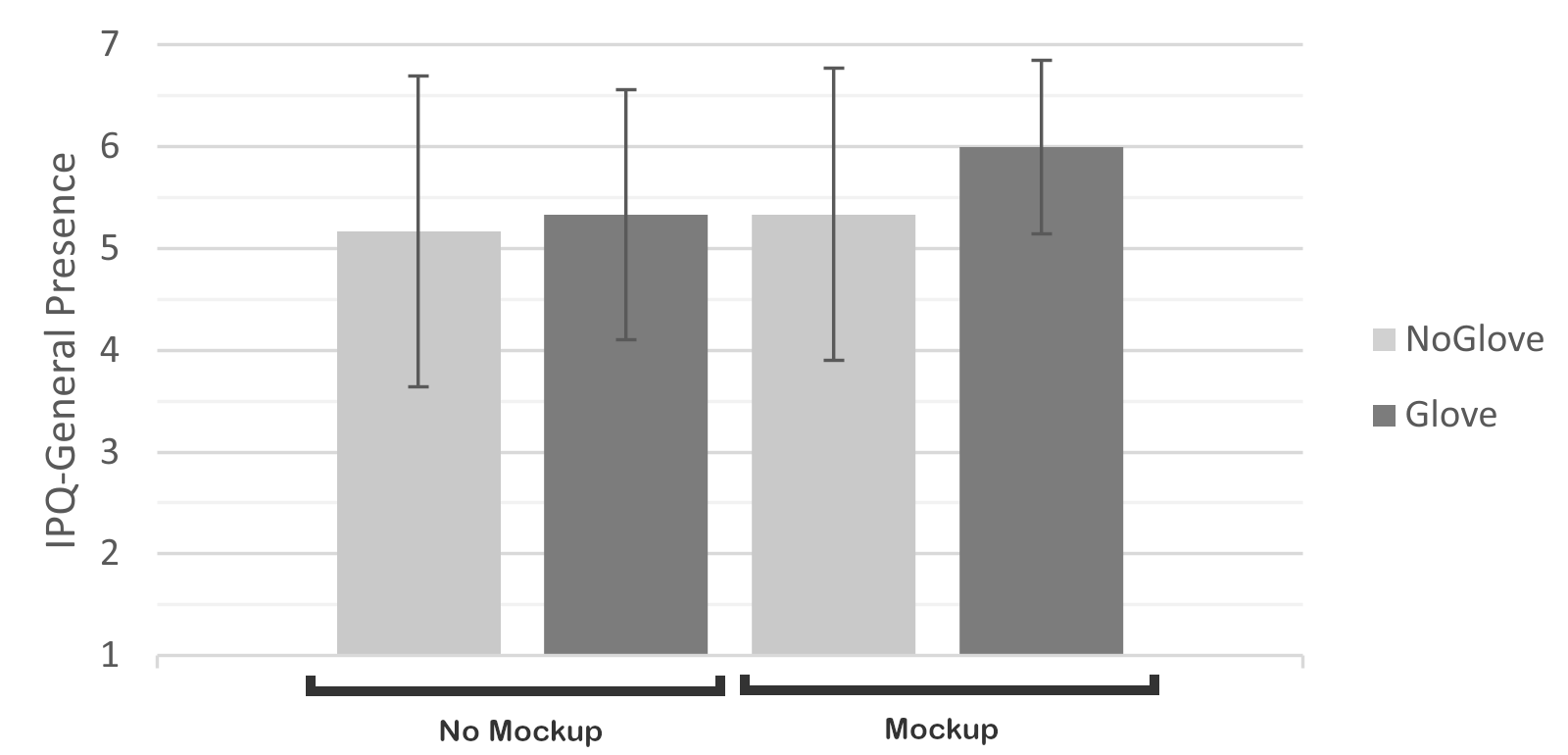}
         \caption{IPQ: general presence}
         \label{Barchart: general presence }
     \end{subfigure}
     \begin{subfigure}[b]{0.42\linewidth}
         \centering
         \includegraphics[width=\textwidth]{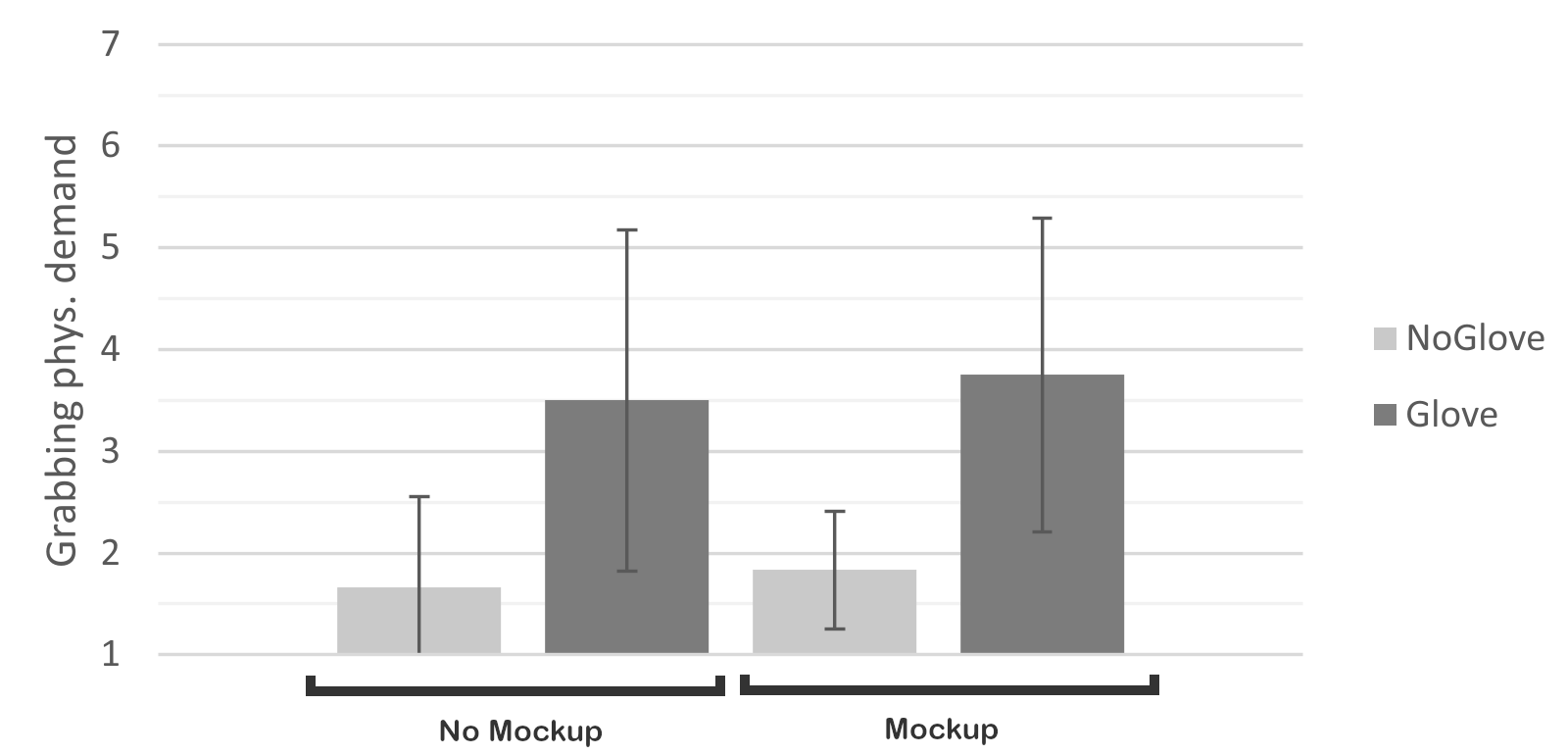}
         \caption{TLX: grabbing task physical demand}
         \label{Barchart: TLX grabbing PD }
     \end{subfigure}
     \begin{subfigure}[b]{0.42\linewidth}
         \centering
         \includegraphics[width=\textwidth]{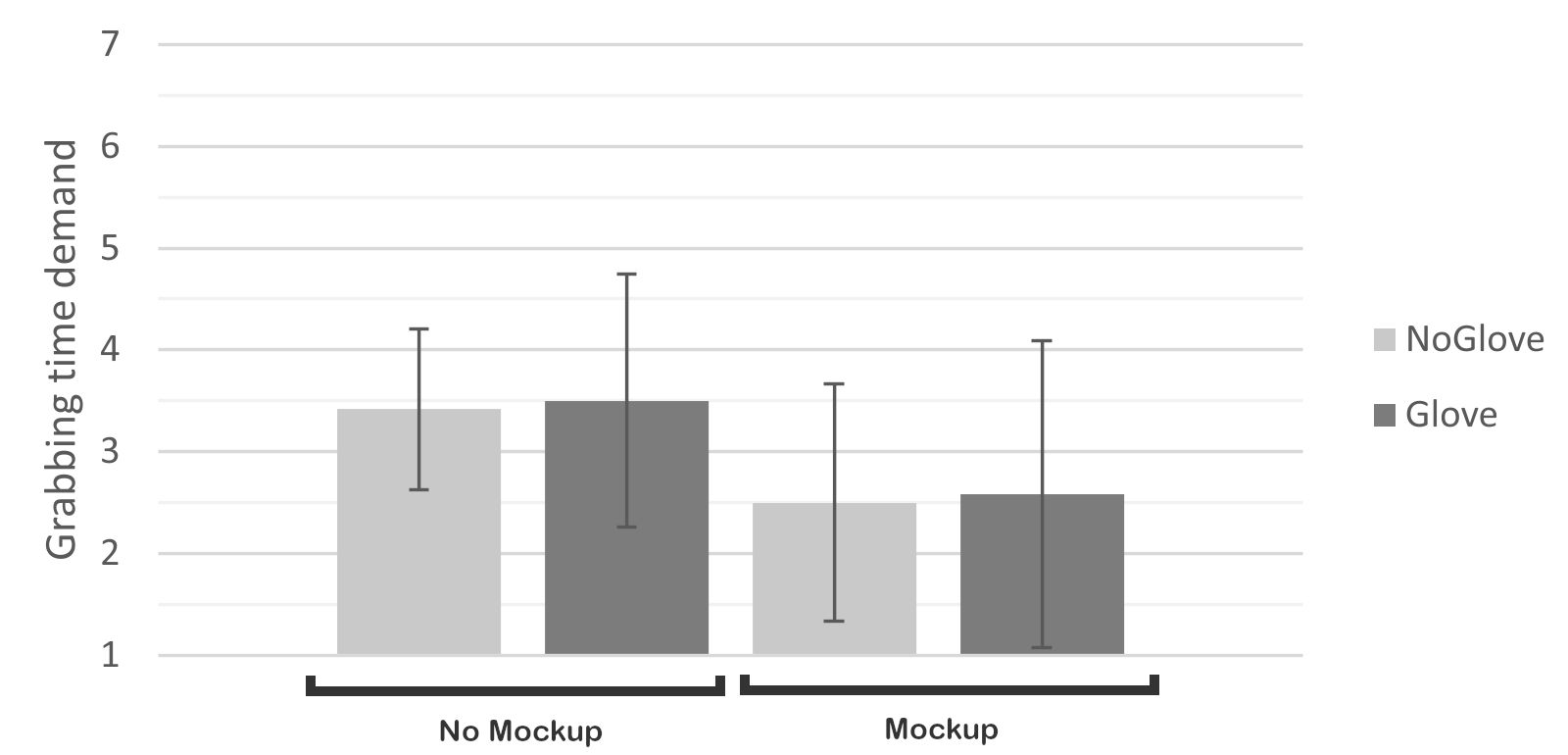}
         \caption{TLX: grabbing task time demand}
         \label{Barchart: TLX grabbing time demand }
     \end{subfigure}
     \begin{subfigure}[b]{0.42\linewidth}
         \centering
         \includegraphics[width=\textwidth]{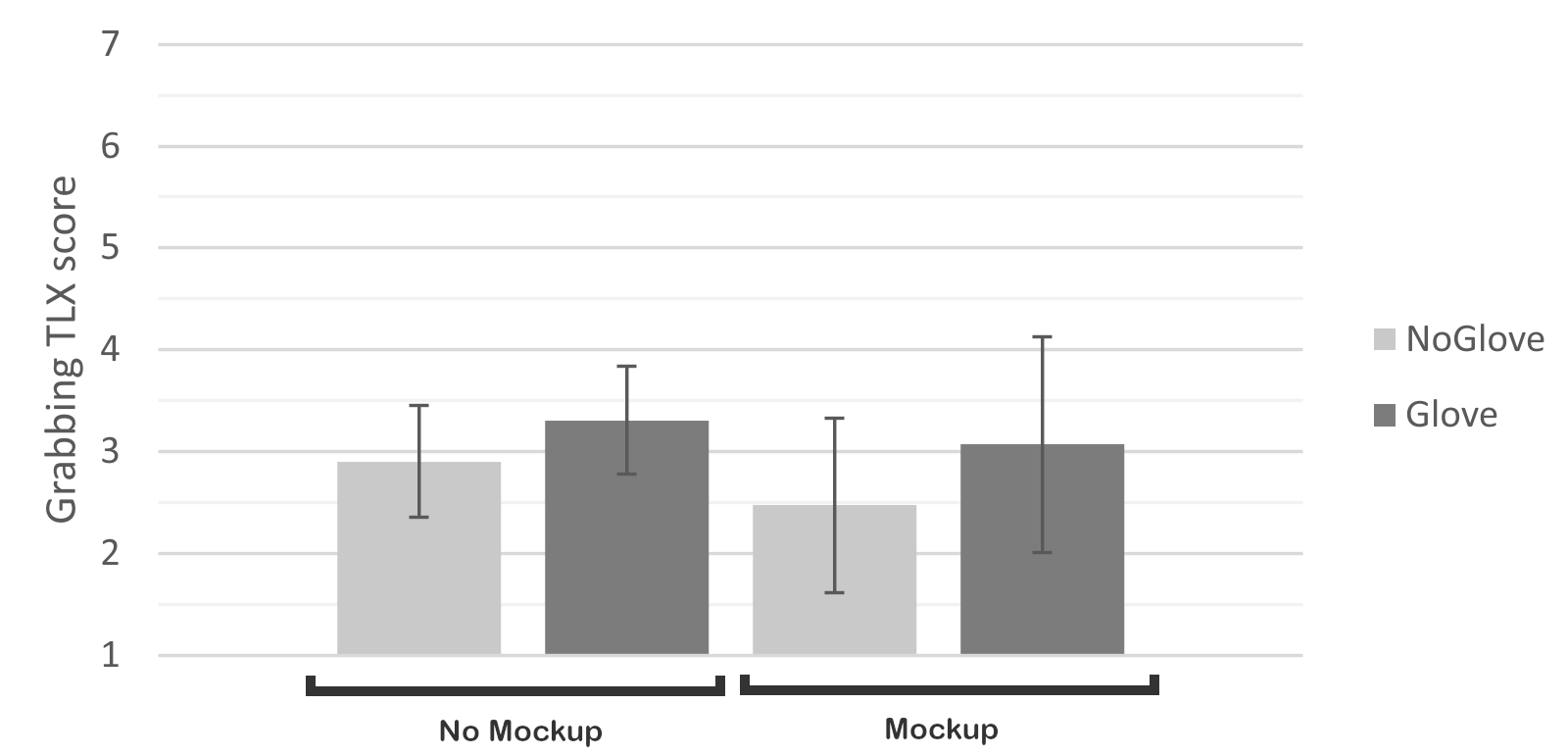}
         \caption{TLX: grabbing task score}
         \label{Barchart: TLX grabbing }
     \end{subfigure}
     \begin{subfigure}[b]{0.42\linewidth}
         \centering
         \includegraphics[width=\textwidth]{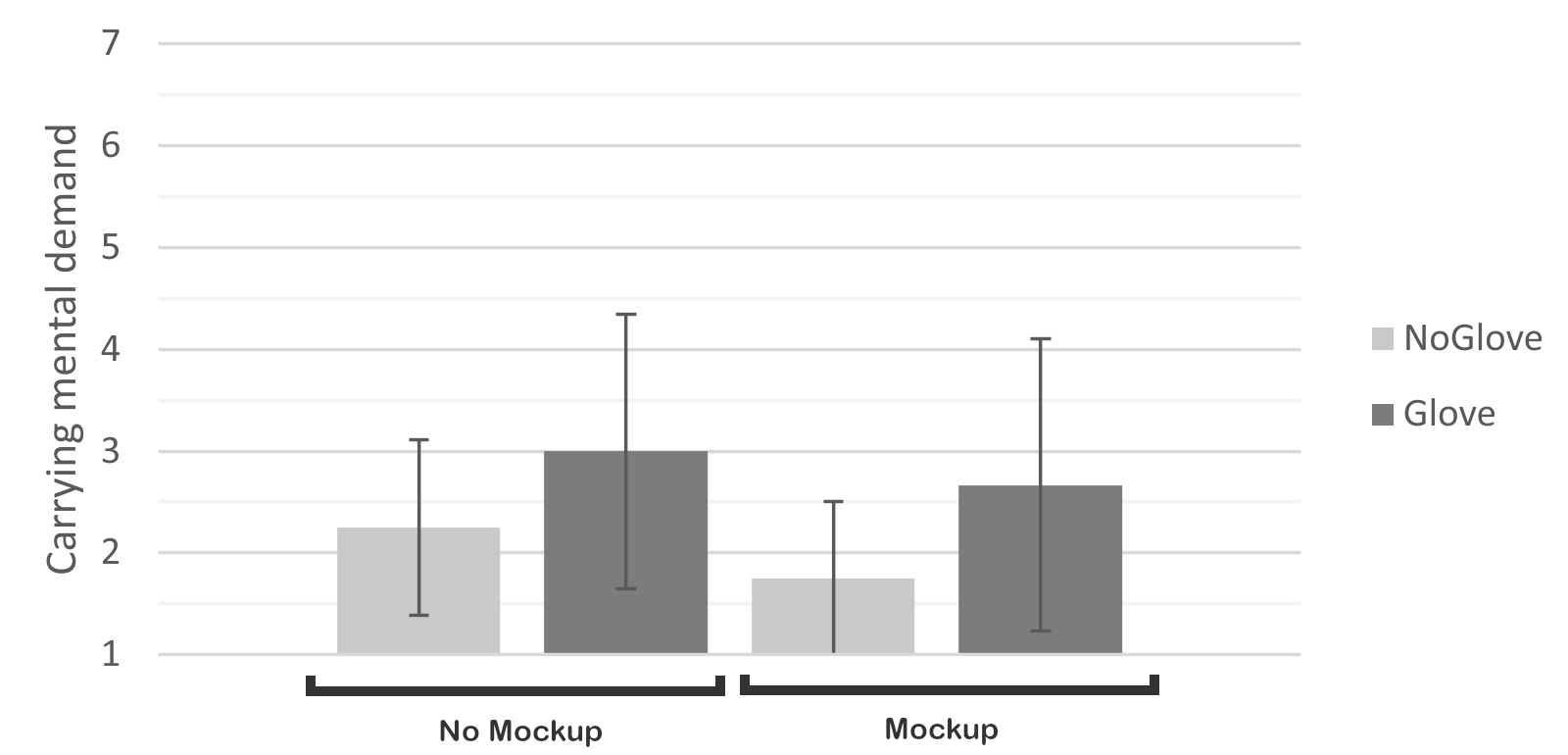}
         \caption{TLX: carrying task mental demand}
         \label{Barchart: TLX carrying MD }
     \end{subfigure}
     \begin{subfigure}[b]{0.42\linewidth}
         \centering
         \includegraphics[width=\textwidth]{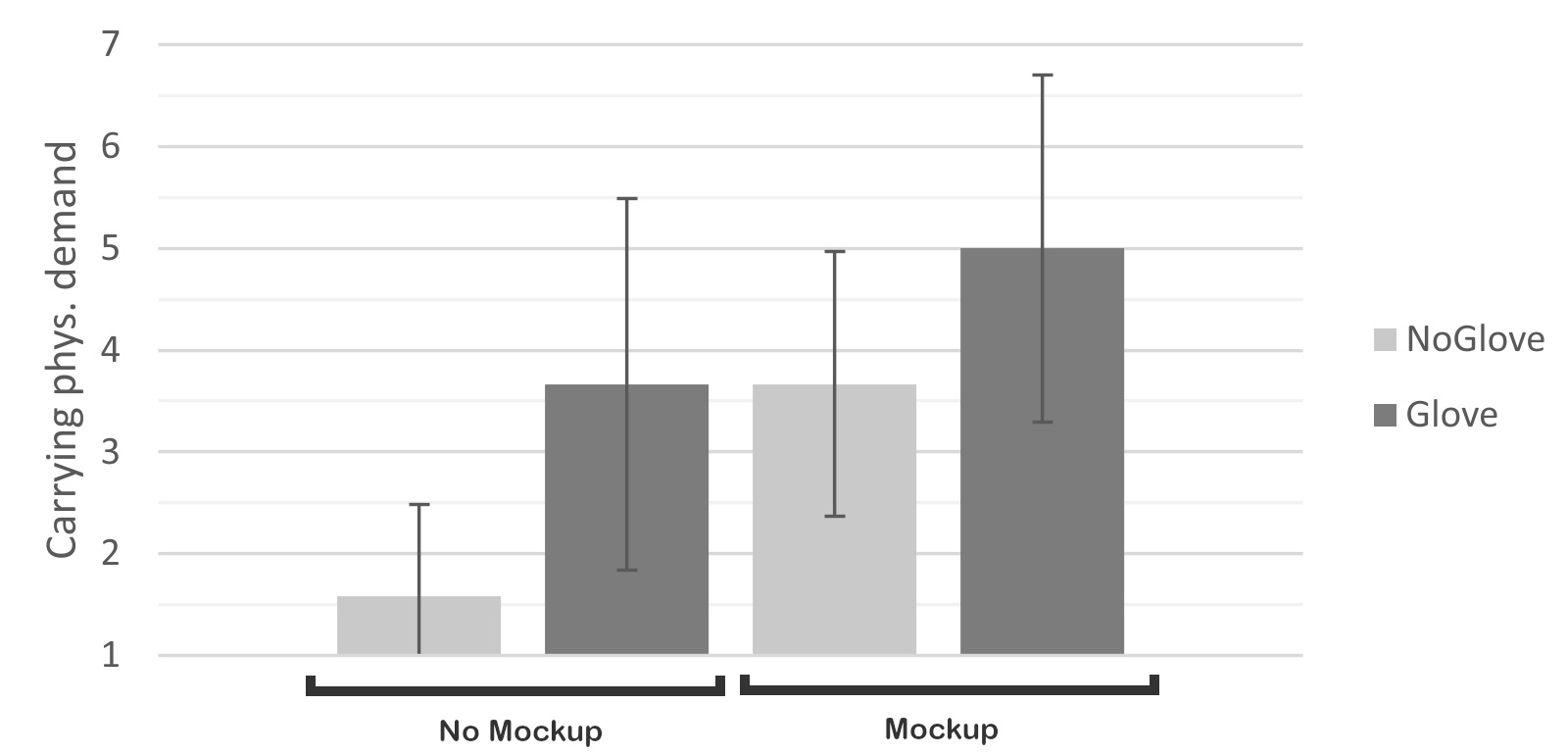}
         \caption{TLX: carrying task phys. demand}
         \label{Barchart: TLX carrying PD }
     \end{subfigure}
     \begin{subfigure}[b]{0.42\linewidth}
         \centering
         \includegraphics[width=\textwidth]{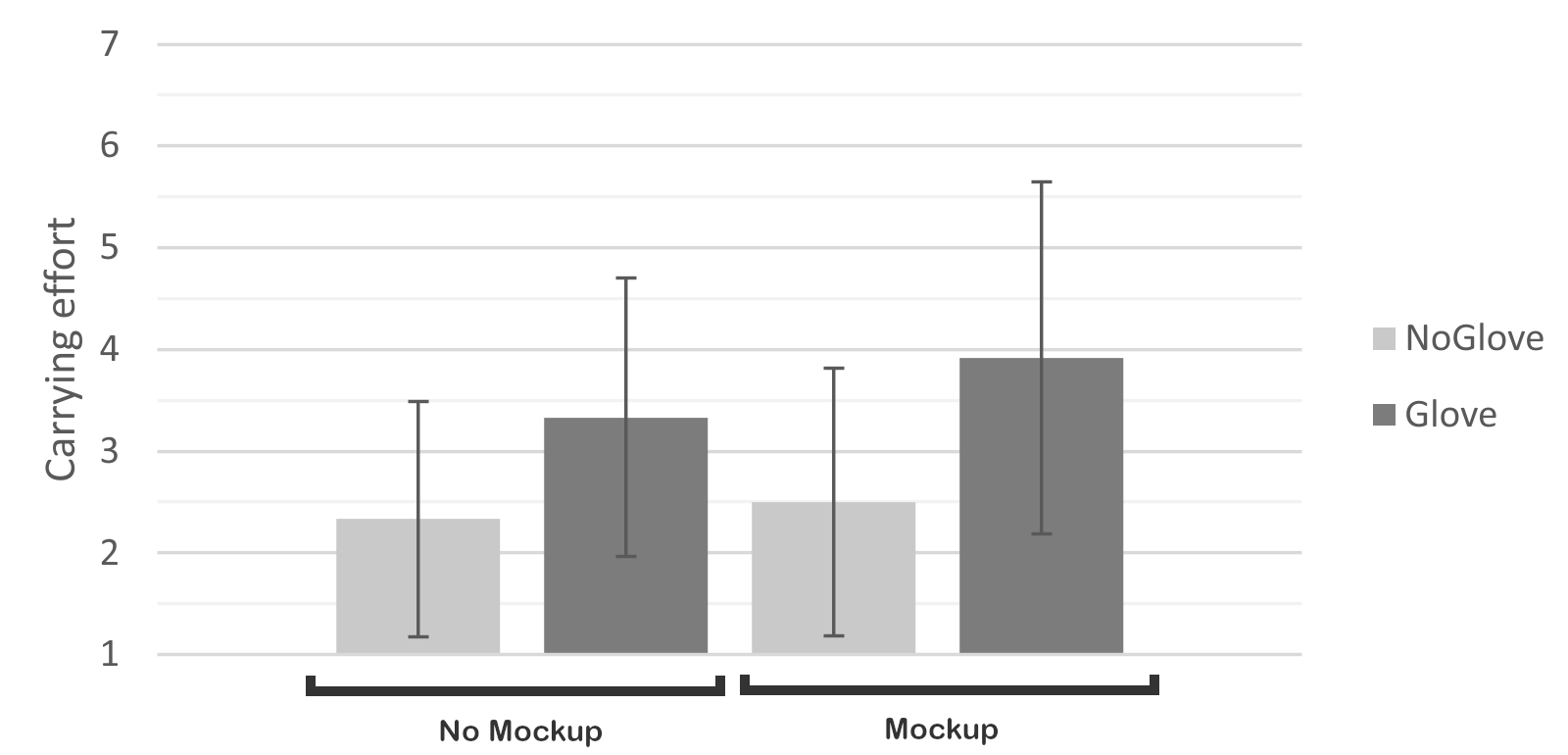}
         \caption{TLX: carrying task effort}
         \label{Barchart: TLX carrying E}
     \end{subfigure}
     \begin{subfigure}[b]{0.42\linewidth}
         \centering
         \includegraphics[width=\textwidth]{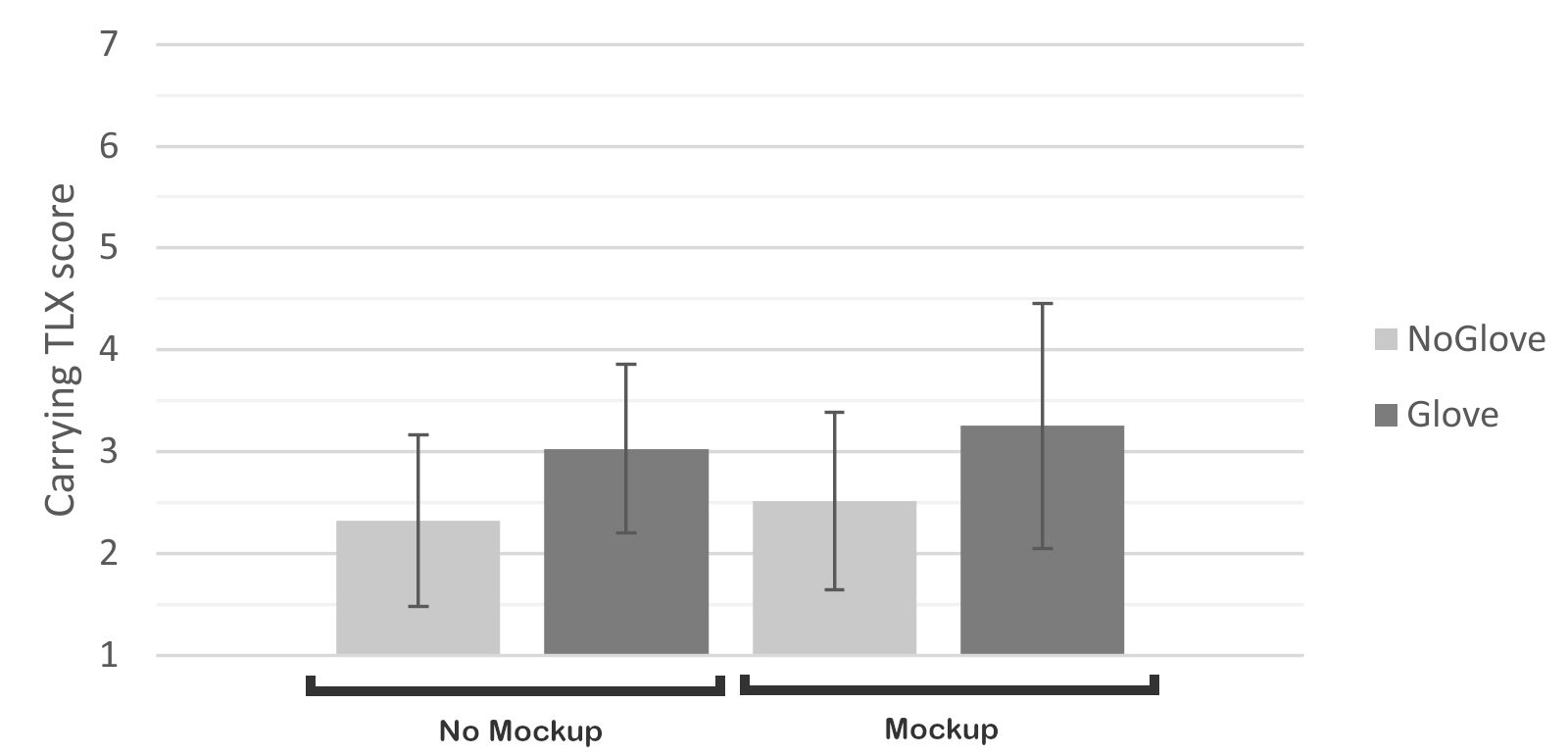}
         \caption{TLX: carrying task score}
         \label{Barchart: TLX carrying}
     \end{subfigure}
        \caption{Barcharts for dependant variables whose results show significant differences.}
        \label{fig:barcharts}
        \Description{The barcharts for the dependent variables whose results show significant differences between groups, namely body ownership, realness, general presence, grabbing task physical demand, time demand and global workload, barbell carrying task mental demand, physical demand, effort, and global workload.}
\end{figure*}

Data were collected through questionnaires, tested for normality (Shapiro-Wilk test), and for equality of variances (Levene’s test) using a threshold value for $p$ of $.05$. For some of the groups, either the normality check or the equality of variances were not verified. However, the \gls*{anova} models are claimed to be robust to non-normally distributed samples and to heterogeneity of variances, but for the latter only when the sample sizes are equal \citep{Pallant2020SPSS}. Sample sizes are strictly the same here with 12 participants per group. Furthermore, none of these assumptions are broken simultaneously. All the results presented in this section come from between-within subjects \gls*{anova} tests and differences are considered significant for $p<.05$.

\subsubsection{Embodiment}
No effects were observed on the sense of agency. Concerning body ownership, there was neither an interaction effect between the gloves and mockup ($F(1, 22)=2.76$, $p=.111$) nor a main effect of the mockup ($F(1, 22)=.080$, $p=.780$). However, there was a significant main effect of the gloves with a large effect size ($F(1, 22)=4.83$, $p=.039$, $\eta_{\text{p}}^{2}=.180$). Descriptive statistics indicated that the average body ownership scores were higher in the gloves condition (see \autoref{tab: DV descriptive statistics}, \autoref{subfig: ownership plot}, \autoref{Barchart: ownership}).

\subsubsection{Presence}
No effects were observed on spatial presence. However, concerning realness, while the analysis revealed no main effect of the mockup ($F(1, 22)=2.27$, $p=.146$), a significant main effect of the gloves was observed with a large effect size ($F(1, 22)=5.54$, $p=.028$, $\eta_{\text{p}}^{2}=.201$), as well as a significant interaction effect between the gloves and mockup with a large effect size (($F(1, 22)=7.54$, $p=.012$, $\eta_{\text{p}}^{2}=.255$)). Descriptive statistics indicated that the average perceived realness score for the group with mockup ($M=4.19$, $SD=.820$) was above the group without mockup ($M=3.40$, $SD=.661$) in the gloves condition, and that the average realness score was higher in the gloves condition (see \autoref{tab: DV descriptive statistics}, \autoref{subfig: Realness plot}, \autoref{Barchart: Realness}).

Concerning general presence, the analysis revealed neither a main effect of the mockup ($F(1, 22)=.707$, $p=.410$) nor an interaction effect between the mockup and glove variables ($F(1, 22)=2.02$, $p=.169$). However, it revealed a significant main effect of the gloves on general presence with a large effect size ($F(1, 22)=5.61$, $p=.027$, $\eta_{\text{p}}^{2}=.203$). Descriptive statistics indicated that the average perceived general presence score was higher in the gloves condition (see \autoref{tab: DV descriptive statistics}, \autoref{subfig: general presence plot}, \autoref{Barchart: general presence }). 

\subsubsection{System Usability}

No effects were observed on system usability. Therefore, every configurations presented a similar level of usability.

\subsubsection{Grabbing task loads}
\gls*{nasa} \gls*{tlx} was used to assess the work load induced by the task of grabbing virtual objects in the different \gls*{vr} configurations. In line with the discussion of \citet{Hart2006TLX}, we chose to compute the global workload score by calculating a mean score of all dimensions. A significant main effect of the gloves on workload with a large effect size was observed ($F(1, 22)=11.2$, $p=.003$, $\eta_{\text{p}}^{2}=.337$). Descriptive statistics indicated a higher workload for the object grabbing task in the gloves condition (see \autoref{subfig: tlx grabbing}, \autoref{tab: DV descriptive statistics}, \autoref{Barchart: TLX grabbing }). 

Among the six dimensions of the \gls*{tlx}, the gloves had a significant main effect on physical demand with a large effect size ($F(1, 22)=36.7$, $p<.001$, $\eta_{\text{p}}^{2}=.625$). Unexpectedly, the mockup had a significant main effect on the time demand with a large effect size ($F(1, 22)=5.29$, $p=.031$, $\eta_{\text{p}}^{2}=.194$). Descriptive statistics indicated a higher physical demand with the gloves (see \autoref{tab: DV descriptive statistics}, \autoref{Barchart: TLX grabbing PD }) and a lower temporal demand in the physical mockup condition (see \autoref{tab: DV descriptive statistics}, \autoref{Barchart: TLX grabbing time demand }). No significant effects were observed for the remaining sub-dimensions.

\subsubsection{Barbell carrying task loads}
Concerning the assembled package transportation task, the tests revealed a significant main effect of the gloves on workload with a large effect size ($F(1, 22)=16.1$, $p<.001$, $\eta_{\text{p}}^{2}=.423$). Descriptive statistics indicated a higher task workload in the gloves condition (see \autoref{subfig: tlx carrying}, \autoref{tab: DV descriptive statistics}, \autoref{Barchart: TLX carrying}). 

Among the six dimensions of the \gls*{tlx}, the gloves had a significant main effect on physical demand ($F(1, 22)=23.5$, $p<.001$, $\eta_{\text{p}}^{2}=.516$), mental demand ($F(1, 22)=10.4$, $p=.004$, $\eta_{\text{p}}^{2}=.322$) and on the perceived effort to achieve performance ($F(1, 22)=17.2$, $p<.001$, $\eta_{\text{p}}^{2}=.438$), all with large effect sizes. Descriptive statistics indicated a higher physical demand, a higher mental demand and a higher effort score in the gloves condition (see \autoref{tab: DV descriptive statistics}, \autoref{Barchart: TLX carrying PD } \autoref{Barchart: TLX carrying MD }, \autoref{Barchart: TLX carrying E}). No other significant effects were observed for the remaining sub-dimensions. The analysis also revealed a significant main effect of the mockup condition on physical demand with a large effect size ($F(1, 22)=12.1$, $p=.002$, $\eta_{\text{p}}^{2}=.356$), and with descriptive statistics showing a higher score in the physical mockup condition (see \autoref{tab: DV descriptive statistics}, \autoref{Barchart: TLX carrying PD }). 

\subsection{Qualitative Analysis}
Our participants displayed varying skill levels, with some requiring additional instructions concerning the assembling procedure, whilst others managed the entire process on their own. Importantly, all participants eventually succeeded in assembling the ALSEP package and transporting it to its destination. Throughout this process, participants were encouraged to comment on any aspect of their experience they found noteworthy. 

The simulation was overall seen as enjoyable, frequently being described as “really cool” (Scientist3, Mockup), “very immersive” (Engineer5, Mockup) and “intuitive” (Engineer6, No mockup). The use of gloves was generally regarded as preferable compared to more traditional interaction interfaces, whilst the physical mockup was often seen as surprisingly heavy. Using both separately or simultaneously appeared to increase perceived body ownership, presence and fidelity of the simulation in potentially meaningful ways, prompting our participants to speculate about the extent to which this may enhance the utility of VR in future ConOps assessments. Below we unpack these reflections in greater detail. 

\subsubsection{Embodiment, presence and fidelity} 
With nearly all of our participants having used traditional VR simulators in the past, a frequently recurring theme was the comparatively “natural” user experience brought about by the use of real gloves. Instructor1 (Mockup) and Engineer5 (Mockup) both argued that grasping virtual objects by clenching one's fist in the real world constitutes a superior interaction metaphor compared to pressing controller buttons followed by the vibration feedback commonly utilised by regular VR interfaces. Manager2 (Mockup) described traditional VR controllers as “awkward”, adding that the intuitive nature of the gloves allowed him to manipulate virtual objects with greater precision than what he could do by moving around hand-controllers. Instructor4 (Mockup) adopted a similar stance, arguing that the sensation of gloves resulted in a greater sense of realism: \\

\textbf{Instructor4 (Mockup):} \textit{“I think it was more realistic because you feel the resistance of the gloves that you wear and you lose the feeling of having a controller in your hand. You have the feeling of not working with controllers but with something that is real.”} \\ 

Correspondingly, the absence of gloves left Instructor4 (Mockup) feeling more self-conscious and aware of interacting with a computer simulation. Similarly, Engineer7 (No mockup) suggested this made the simulation feel more like a video game and Scientist4 (Mockup) argued his experience became “way less realistic” without gloves. 

The effect of physical gloves on perceived body ownership was nevertheless not immediately clear-cut and varied depending on the individual, even within groups. Several participants reported that the absence of gloves caused a discrepancy between the visual feedback (i.e., the virtual gloves they saw in VR) and the tactile feedback (i.e., the physical controllers they were holding in their hands), which in turn may have impeded their sense of virtual body ownership. Instructor2 (No mockup) and Instructor1 (Mockup), for instance, reported feeling able to better see the tasks from the astronaut’s point of view, arguing that gloves would be necessary for a realistic assessment of the simulated operation.

On the other hand, Engineer 6 (No mockup) and Engineer9 (Mockup) both suggested that the visual feedback alone was sufficient to trick them into feeling as if they were wearing real gloves. Conversely, Astronaut1 (Mockup) argued that even while wearing the real gloves, it was still difficult to identify with her virtual avatar, citing the lack of mass and inertia manifested by other objects in the virtual space.

After completing the assembly procedure and engaging with the physical mockup, their perception began to change markedly. Nearly all participants in the physical mockup experimental condition reported that grappling with the barbell mockup significantly improved their ability to put themselves into the shoes of the Apollo 12 astronauts and empathize with their situation. Engineer8 (Mockup), for instance, claimed the sensation of weight helped him feel as though he was conducting a real lunar EVA “actually holding something”, whilst Engineer5 (Mockup) declared he was now able to understand “how much of an athlete an Apollo 12 astronaut had to be”.

The heightened physical exertion induced by the mockup's weight elicited substantial reflections among participants, with all of them agreeing that it enhanced the simulation’s immersion and realism in meaningful ways. As Manager2 (Mockup) explained, traditional audiovisual simulations fail to factor-in crucial aspects of real-world astronaut operations, resulting in experiences that are trivialised and may lack validity:

\textbf{Manager2 (Mockup):} \textit{Feeling the weight is extremely important. Because when you do those kinds of simulations in VR, you're factoring out a lot of elements that during the real-life execution of an activity really make a difference. The weight… the inertia of an object is one such thing. So to have an understanding of not just how big something is visually, but also how heavy it will be… I think it's something that really adds tons of value, quite literally, to the simulation.}

Following the same line of reasoning, Astronaut1 (Mockup) welcomed the presence of mass and inertia, arguing the physical mockup was sufficient to enable assessment of relatively simple movements, such as the lifting up and placing down of the ALSEP barbell. 

The general usefulness of weight perception in the simulated procedure was further underscored by the fact that participants in the virtual mockup experimental condition frequently advocated for the addition of a physical mockup that would lend the virtual ALSEP package a sensation of mass, unbeknownst that this was what participants in the other experimental condition received. 

Upon closer examination, it became evident that the increased physical effort required was not solely due to the weight of the physical mockup. The rubbery gloves, in fact, exacerbated this perceived challenge even further. As illustrated by Instructor1’s (Mockup) and Astronaut2’s (Mockup) comments below, obtaining a secure grip and manipulating the physical mockup became notably more difficult when wearing gloves:\\

\textbf{Instructor1 (Mockup):} \textit{It is more realistic to wear virtual gloves when you also feel that you wear them physically. Grabbing the mockup became more difficult, which is realistic, when I was wearing the [real] gloves. Grabbing and transporting it without the gloves felt much easier.} 

\textbf{Astronaut2 (Mockup):} \textit{I was struggling with the bar that kept slipping out of my hands. The task was not any different from when I was not wearing the gloves. But now it was like… more complicated. And Frustrating.}\\

Astronaut2 (Mockup) was not alone in characterizing the experience of manipulating the mockup while wearing gloves as “frustrating”. Other participants used terms such as “cumbersome” (Instructor4, Mockup) or “quite difficult” (Scientist1, Mockup). However, they all concurred that, in the context of ConOps assessment, inducing such frustration was beneficial. As Manager2 (Mockup) pointed out, experiencing a degree of discomfort is important because it more faithfully replicates the demands inherent in an actual operation. 

Interestingly, some participants noted that the gloves also seemed to enhance the perceived weight of \textit{virtual} objects. Engineer7 (No mockup), in particular, expressed the feeling that wearing gloves conferred a sense of mass to the virtual barbell, making its handling feel more physically demanding. Instructor3 (No mockup) and Engineer6 (No mockup) disagreed, stating that wearing the gloves did not significantly alter their perception of the weight of items in the virtual space. 

All in all, what became clear through the study was the significant synergy between real gloves and the physical mockup. According to Engineer5 (Mockup), using gloves while interacting with the physical mockup felt "more impactful than without them," and Engineer3 (Mockup) emphasized that it's the combination of both elements that "makes things a lot more immersive". 

Beyond enhancing the fidelity and realism of the simulation, our passive haptic interfaces thus appeared to boost the users' overall sense of presence and increase their sense of ownership over their virtual bodies. The manner in which this may affect the efficacy of VR-based ConOps assessments became a focal point for further discussion.

\subsubsection{Implications for ConOps Assessment}
Upon safely returning to Earth, the Apollo 12 astronauts underwent a debriefing procedure, during which they were asked to reflect on their mission experience. Their comments were documented and subsequently made available to the public by NASA’s Mission Operations Branch Flight Crew Division \cite{NASA1968Debriefing}. Additional interviews with the crew were also conducted to help define EVA systems requirements for future missions \cite{Connors1994InterviewsDesign}. 

\begin{table*}[hbt!]
\small
    \centering
    \begin{tabular}{p{.20\textwidth}p{.31\textwidth}p{.31\textwidth}}
    \hline
        \textbf{Element} & \textbf{Apollo Feedback Compilation} & \textbf{Quote} \\ \hline
        Placing objects & Keeping objects in place on the lunar surface was challenging due to reduced gravity, with dust contamination being brought up as a recurring issue. & \textit{"We got back to the ALSEP and started a normal deployment.[...] as soon as we put the packages down on the surface, they began to accumulate dust"} (Charles Conrad Jr., Apollo 12, 1969). \\ \hline
        Gloves & Holding the grip on tools was inducing some muscle fatigue as astronauts needed to exert constant pressure. & \textit{"the gloves don't want to stay closed".}(Alan Bean, Apollo 12, 1969) \\ \hline
        Barbell carrying task & It was hard to carry packages due to their weights, restricted body movements and the gloves that were not really closing and thus causing a poor grip. & \textit{"The workload carrying it out was just about the same as I had guessed from working on Earth. The hard part is holding that weight in your hands. [...] the combination of the weight, the fact you're moving along, and the fact that the gloves don't want to stay closed tends to make it fairly difficult task."} (Alan Bean, Apollo 12, 1969) \\ \hline
        Muscle tiredness & Tiredness was mainly felt through the hands and arms for the barbel carrying task. & \textit{"It's not your legs that get tire; it's a combination of your hands and arms and it just makes you tired."} (Alan Bean, Apollo 12, 1969)\\ \hline
        ConOps improvements & The crew advised that for long distances, some technical aids should be considered rather than hand carrying. & \textit{"I would say it would be acceptable to carry it this way for distances up to 500 feet. You will want to have a strap that fits over your shoulder, or something like that."} (Alan Bean, Apollo 12, 1969) \\ \hline
    \end{tabular}
    \caption{Compilation of the Apollo crew feedback \cite{NASA1968Debriefing, Connors1994InterviewsDesign}}
    \label{tab: Apollo feedback}
    \Description{The compilation of the Apollo crew feedback, categorised by themes of interest.}
\end{table*}

By subjecting this body of work to a qualitative thematic analysis, we were able to identify several challenges that astronauts encountered during the ALSEP package deployment. Table \ref{tab: Apollo feedback} presents a compilation of the primary problem areas related to assembling, manipulating and relocating  ALSEP components, each accompanied by a relevant astronaut comment for illustration. As detailed in the remainder of this section, the Apollo 12 astronaut feedback appeared to be largely congruent with the qualitative reflections made by our study participants, revealing some notable overlaps. 

Although our simulation only reenacted the reduced lunar gravity visually, it succeeded in eliciting relevant reflections. Several participants were cautious to prevent colliding with equipment they had positioned on the lunar surface. Instructor4 (Mockup) and Engineer4 (No mockup) explicitly credited the simulated lunar gravity with making it easier to accidentally knock over objects. Similarly, drawing on their expertise, many participants also recognized the risk of regolith (moondust) contaminating vital equipment. Upon placing the assembled mockup on the ground, Manager2 (Mockup), for instance, noted it makes him feel “really uncomfortable to see this thing [the \gls*{rtg}] in direct contact with the regolith.” 

As expected, our haptic interfaces proved to play a greater role in shaping participants’ reflections concerning the grabbing and manipulation of objects. The stiffness of the rubber gloves, combined with the positioning of the controller on the exterior of each glove, led most participants to experience a heightened sense of exertion when attempting to grasp virtual objects: \\ 

\textbf{Instructor2 (No mockup):} \textit{“The controller design means you need to keep pressing down the button through the gloves. So that's of course more effort.” }\\ 

These efforts increased further when participants interacted with the physical barbell mockup. Engineer3 (Mockup) and Scientist1 (Mockup) both experienced significant challenges in achieving a secure grip. Instructor4 (Mockup) argued the gloves required him to exert “more grip strength”. As Scientist4 (Mockup) explained, the stiffness of the gloves was a major contributing factor to these difficulties: \\ 

\textbf{Scientist4 (Mockup):} \textit{“With gloves I had to have my hands much more open. So you have to use more strength to grab it [the mockup]. You can say that it felt heavier because I couldn't fully close my hands. So you have to use more strength, it becomes a physical experience.”} \\ 

Several participants needed to place the mockup on the ground to take a break during the transportation task before continuing. Manager2 (Mockup) pointed out that the combination of gloves and the mockup increased the physical strain “for the hands”. Scientist3 (Mockup) and Engineer3 (Mockup) both also mentioned feeling hand fatigue. Similarly, after finishing the transportation task, Engineer5 (Mockup) humorously noted that he "shouldn't have done those deadlifts in the gym earlier."

The impact of the physical mockup on participants' perspectives also became evident when we inquired about potential enhancements to the simulated ConOps. Much like the Apollo 12 astronauts before them, participants who interacted with the physical mockup were primarily focussing on lessening the workload associated with the barbell transportation task. For instance, Engineer9 (Mockup) proposed a redesign of the ALSEP package to enable two astronauts to carry it simultaneously. Scientist4 (Mockup) suggested providing astronauts with a cart for transporting heavy equipment, while Manager2 (Mockup) recommended mounting wheels directly on the assembled package.

Conversely, participants who engaged with the virtual barbell mockup tended to focus more on improving different facets of the assembly process. Scientist5 (No mockup), for instance, proposed simplifying the construction of the carry mast by replacing its two components with a single telescopic tube. Instructor2 (No mockup) suggested adding more handles to the individual components for easier manipulation, while Engineer6 (No mockup) recommended labeling the components to highlight crucial attachment points.

In general, while the haptic interfaces did seem to help elicit more valid reflections from our participants, several issues that affected the simulation's fidelity also came to light. In particular, the absence of full-body suit constraints drew significant criticism, with Astronaut1 (Mockup) arguing this oversight had downplayed the complexity of many actions during the simulated operation. She concluded that our VR system, in its current state, could be valuable for "an initial assessment of astronaut choreography" but additional work would be necessary before it could be utilized for evaluation of more complex procedures. 

Given such limitations, it is important to approach the comparison between our findings and an authentic lunar mission with caution. Nevertheless, when considered in conjunction with the quantitative and qualitative results presented throughout this paper, the apparent similarities between our participant’s comments and the feedback provided by the Apollo 12 astronauts can be seen as indicative of a general tendency towards alignment between virtual and real experiences brought about by haptic interfaces. The significance of this will be discussed in detail in the following section.

\section{Discussion}\label{section: discussion}

This study aimed at investigating the impact of different passive haptic modalities on the senses of presence, embodiment and workload in the context of VR-based assessments of \gls*{conops} for future lunar surface activities. Using astronaut gloves and a low-fidelity mockup of a science experiment package from the Apollo era, we produced 4 different \gls*{vr} configurations that were tested in a mixed-design experiment by a panel of 24 experts specialising in aerospace engineering and human spaceflight. Following a within-subjects study design, each participant completed the VR simulation twice: once while wearing gloves and once without them. Following a between-subjects study design, half of the participants used a physical mockup, while the other half experienced its virtual counterpart. Statistical analyses showed that wearing gloves resulted in significantly higher body ownership, general presence and realness of the \gls*{ive}. Furthermore, interactions between wearing gloves and using the physical mockup resulted in significantly higher scores in terms of realness of the virtual environment. These results, along with the qualitative evaluations, are discussed in this section in light of related studies. 

To begin with, the participants were asked to report on the general usability of the interface through the \gls*{sus} \cite{Brooke1996SUS}. No significant differences between the distinct interface configurations were observed. Mean usability scores of all configurations ranged from $M=62.1$ to $M=66.2$ out of 100 (\autoref{tab: DV descriptive statistics}), which put them in the acceptable range of systems \cite{Bangor2008SUS}. This demonstrates that all proposed \gls*{vr} setups can provide a similar level of usability. Depending on the context, both the physical interfaces and their virtual counterparts can be considered for \gls{conops}-assessment purposes.

\subsection{Embodiment}
As expected, our quantitative results revealed an influence of wearing gloves on the sense of body ownership, thus validating our first hypothesis (\textbf{H1}). This expands on previous results from \citet{Gorisse2023Wearable} who observed no significant impact of wearing a physical wristband while experiencing it in an \gls*{ive}. It can then be supposed that leveraging interactivity of the wearable, namely gloves that fundamentally differ from a wristband as they foster the user's prehension, may have contributed to experiencing greater body ownership. Even if qualitative findings showed conflicting opinions among participants regarding the effects of wearing gloves on embodiment, it appears that the latters greatly contributed to having our participants experiencing enhanced ownership toward the virtual body.

However, results revealed no significant interaction between wearing the gloves and carrying the physical mockup on body ownership. Despite participants praised the haptic sensations and challenge afforded by using gloves in conjunction with the mockup, it appears likely that this effect solely originates from wearing gloves. Also, as expected, there was no impact of the barbell mockup alone on body ownership. None of the haptic system influenced the sense of agency of the participants, suggesting that unlike with a grabbing and reaching task \cite{Medeiros2023BenefitsPassiveHaptics}, introducing tactile sensations with passive props for a grabbing task does not increase agency. All in all, actual astronaut gloves contributed to an improved perceived fidelity of the simulation by narrowing the gap between astronauts' physical experience and participants taking part in the simulated \gls*{conops}.

\subsection{Presence}
Perceived presence in the virtual environment was assessed based on the general presence (i.e. "being there"), spatial presence (i.e. feeling physically present), and realness (i.e. perceived realism) dimensions measured through the \gls*{ipq} \cite{Schubert2003IPQ}. As expected, the gloves positively influenced the sense of general presence in, and realness of the \gls*{ive}. This aligns with previous work showing that passive haptics increase the sense of presence in \glspl*{ive} \cite{Insko:2001:PHF, VicianaAbad2010HapticInteraction}. However, the physical mockup did not influence any of the dimensions of presence, which contrasts with the aforementioned previous findings. Consequently, \textbf{H2.1} was partially verified and \textbf{H2.2} was disproved.

Wearing gloves while manipulating the physical mockup, also positively impacted the perceived realness of the virtual environment. In turn, the interaction between the two haptic interfaces caused the virtual environment to be perceived as more authentic. This finding corroborates and expands on existing work concerned with passive haptic modalities in VR.  \citet{VicianaAbad2010HapticInteraction}, for instance,  demonstrated that using a physical pen in \gls*{vr} without surface collision feedback was rated worse in terms of presence than the condition with collision surface. By cautious analogy, we may argue that carrying the physical mockup would provide a better sensation of surface and weight feedback than the hand controllers alone would do. It therefore enhances the impact of the mockup and increases the sense of presence through realness of the interaction. This assertion is corroborated by the feedback provided by our expert participants, such as Instructor1, Manager2, and Engineers 3 \& 5, who all praised the realness of the physical interaction with the mockup.

\subsection{Validity for ConOps Assessment}
In addition to tracking the role of embodiment and presence in shaping the overall fidelity of \gls*{vr} simulations, we also measured perceived task load and gathered relevant user feedback. Through these efforts, we were able to conduct an initial evaluation of the feasibility of using \gls*{vr} setups enhanced by passive haptics for the purpose of \gls*{conops} assessments. Workload assessment of the object grabbing task revealed a significant negative impact of wearing gloves. We therefore validate \textbf{H3.1} and reject \textbf{H3.2}. More specifically, the gloves induced a higher physical demand, as illustrated by our participants' concerns: introducing haptic stimulation through gloves resulted in raising considerations among experts that strongly align with the actual Apollo 12 crew feedback, which stated that the gloves and barbell design made the transportation task relatively difficult. On another note, the physical mockup lowered the perceived time demand during the grabbing task. Indeed, some participants reported that grabbing the physical mockup felt more natural, and may in turn be faster than apprehending the virtual barbell grabbing. It must be mentioned that many participants reported the snapping-like grabbing interaction metaphor was not natural and thus contributed to a degradation of their scenario experience, further highlighting the value of tangible props.

Concerning the task of carrying the barbell to the drop-off point, tests exclusively revealed that the gloves induced a greater workload, therefore validating \textbf{H4.1} and rejecting \textbf{H4.2}. More specifically, wearing gloves increased participants' mental load, physical load and effort needed to achieve the goal. Participants indeed mentioned that wearing gloves reduced their grip which correlates with the Apollo crew feedback. However, this result is surprising given that it also applies to participants wearing gloves but without the physical mockup. Indeed, some participants reported that this task required them to hold the grip continuously because of the controllers' button requiring to be pressed, which proved rather challenging with the gloves and the bar controllers in the condition without the physical mockup. This statement also strongly corroborates the feedback from Apollo crews. Thus, combination of the gloves and controllers, that required to hold their button tight, was sufficient to bring the perceived fidelity of this task closer to what it was during the actual Apollo mission and may explain small differences between groups. Finally, as expected, using the physical mockup increased the physical demand that participants experienced during the transportation task of the assembled package. Our panel reported they could better empathise with the astronauts that had to carry this weight and were therefore able to propose some transportation improvements. All in all, these findings suggest that passive haptics constitute a step forward in terms of facilitating valid evaluations of \gls*{conops} through \gls*{vr}. Additionally, the sliding-like navigation metaphor during the barbell transportation phase was described as unnatural by the experts. This choice was made to cope with the limitations inherent to \gls*{vr} regarding locomotion in limited spaces. Such drawbacks could be addressed by leveraging larger physical surroundings that allow for natural or redirected walking~\cite{Razzaque:2001:RedirectedWalking, Nilsson:2018:RDWReview, Simeone:2020:SpaceBender} with the mockup, leading then to a more accurate evaluation of transportation tasks.

\section{Limitations and Future Work}\label{section: limitations}



Significant findings were made with regards to improving perceived fidelity of \gls*{vr}-based \gls*{conops} evaluations in the context of lunar surface missions. Given the unusual use case of this study, we would advise exercising caution when attempting to apply these findings to other contexts. Further research is required to explore the cross-domain applicability of the principles elaborated above. Nevertheless, qualitative findings may benefit not only to space-oriented \gls*{conops}, and psychometric results were obtained with a demographically diverse panel and should therefore benefit designers of \gls*{vr} setups using passive haptic props in domains beyond human spaceflight. Recreating the extensively documented Apollo 12 mission limited us to using outdated hardware like the barbell. But more than evaluating the design itself, the aim was to assess whether or not \gls*{vr} used in synergy with passive haptic proxies could raise critical reflections about the hardware design and associated role in the \gls*{conops}. For this purpose, a design like the Apollo barbell for which drawbacks were known was deemed suitable. \gls*{vr} is already used to evaluate future spacecraft and surface hardware designs \cite{Nilsson2022EL3, Nilsson2023ShapeHumanity}, but further research could investigate its use in combination with haptics to assess upcoming mission \gls*{conops} making use of tailored contemporary equipment. Another concern arose about the astronaut gloves we used as they were unpressurised, making the experience easier than for the actual Apollo crew. However on-going design endeavours focus on improving glove ergonomy for the Artemis program. We might therefore speculate our glove mockup formed a middle ground between the old and upcoming new designs.  Finally, it must be mentioned that unlike the Apollo program, future missions will likely target the lunar south Pole. This area presents harsher environmental conditions, like atypical lighting causing long pitch black shadows, that will make assembly tasks and surface operations even more challenging. Despite such future missions parameters were not considered in this study, we would argue \gls*{vr} is uniquely positioned to simulate these elements.

With regards to \gls*{vr}-based lunar surface simulations, we would suggest further research to explore the following directions: some of our participants expressed the desire of having a representation of the movement of the fingers while some others mentioned that haptic feedback could be embedded in the glove, therefore investigating the use of a built-in solution combining finger tracking and haptics in a single glove could benefit the community. To address the raised lack of haptically perceivable weight during \gls*{vr} simulations, it might be worthwhile to explore combinations of the utilised haptic interfaces (i.e., gloves or mockups) with mixed feedback concepts or software-based illusion techniques~\cite{Zenner:2022:AdvancingProxyBasedHapticsInVR}. To add varying sensations of weight, for example, the concept of weight-shifting~\cite{Zenner:2017:Shifty} or drag-changing~\cite{Zenner:2019:Dragon} feedback could be adapted when designing simulation gloves or equipment mockups. Alternatively, or as an addition, pseudo-haptic algorithms~\cite{Lecuyer2009SimulatingFeedback, Samad:2019:PseudoHapticWeight} and hand redirection techniques~\cite{Azmandian:2016:Retargeting, Zenner:2021:DPHFHR} could be employed to augment perceptions of weight, inertia, and balance during lunar simulations. 
Also our participants raised that having a full virtual body representation of the \gls*{eva} suit, that would physically respond to events in \gls*{ive}s, could improve their reflection, especially if the constraints induced by the actual suit are emulated. Such emulations could potentially be realised also through illusion-based techniques like hand redirection~\cite{Zenner:2019:HandRedirectionThresholds} if wearable suit mockups are not available. On another note, it could be valuable to further increase immersion, and therefore the environmental fidelity, by combining \gls*{vr} with physical testbeds that more accurately replicate some lunar surface features. Finally, proposing a one-to-one comparison between a \gls*{vr}-based simulation of a scenario and the same operations that would take place in the aforementioned testbeds would greatly contribute to increasing knowledge about the value of \gls*{vr}-based \gls*{conops} assessment in comparison to analogue field studies.
\section{Conclusion}\label{section: conclusion}
The presented study aimed at investigating the impact of two passive haptic interfaces, coupled with VR simulations, on important dimensions of the user experience in \gls*{vr}-based \gls*{conops} assessments for lunar exploration. The senses of presence and embodiment were considered along with an evaluation of lunar surface \gls*{conops}. Experts in the fields of human spaceflight were immersed in a partial simulation of the Apollo 12 surface mission. Half of them put their hands on a physical mockup inspired by the Apollo 12 equipment as a between-subjects variable. In addition, they were all able to experience the scenario twice, once with, and once without wearing authentic astronaut gloves as a within-subjects variable (\autoref{fig:collage}). The tests revealed significant effects of the passive haptic proxies: gloves on their own increased the sense of embodiment through body ownership, and of general presence along with perceived realness of the situation; wearing gloves in combination of using the barbell mockup resulted in significantly higher perceived presence through realness as well. The observed dependent variables demonstrate an enrichment of the fidelity of the VR scenario with respect to the actual simulation, which in turn allowed experts to point out concerns about this \gls*{conops} that strongly corroborate actual Apollo12 astronauts feedback. These results therefore suggest that increasing fidelity of the simulation through increasing embodiment and presence by leveraging the use of passive haptics may contribute to a greater validity of VR-based \gls*{conops} assessments.

Taken together, these results contribute to the growing body of knowledge on the impact of passive haptics on embodiment and presence. The propensity of these senses to prompt participants to act more as if the simulation were real and to place greater emphasis on the astronaut condition induced an increase in simulation fidelity. This is the reason why we argue that haptics, and consequently embodiment and presence, are paramount and should be leveraged by \gls*{vr}-based \gls*{conops} review applications designers.

\begin{acks}
We extend our heartfelt gratitude to the experts, trainers, scientists and students who took part in our study, and especially to the astronauts: Mrs Samantha Cristoforetti, Mr Matthias Maurer and Mr John McFall. Their invaluable insights, dedication, and commitment significantly enriched our research. Their willingness to share their experiences and expertise allowed us to gain a deeper understanding of the benefits human-computer interactions could bring beyond our terrestrial bounds.

We also warmly thank Mrs Lora-Line Faure for participating in the recording of the video figure accompanying this publication.
\end{acks}

\bibliographystyle{ACM-Reference-Format}
\bibliography{references}

\appendix
\section{Set of questions to lead the discussion} \label{Appendix: think aloud}
These questions served as a basis for the experimenter to guide the participant's reflection following the think-aloud approach.\\

"I will invite you to reflect on this VR setup, specifically I'm interested in understanding its potential use to enable evaluation of key elements of the Apollo 12 mission."\\

Grabbing and placing objects:
\begin{itemize}
    \item Given the VR setup and hardware, what is your experience grabbing and placing objects in the virtual environment?
    \item Could you speculate what the experience might be like for astronauts doing this task on the Moon?\\
\end{itemize}

Handling and carrying the assembled packages (barbell):
\begin{itemize}
    \item Do you think the VR and hardware setup used to simulate carrying the set of equipment help you infer what it would actually be like to carry out the same procedure on the actual Moon?
    \item Could you then reflect on an astronaut carrying the set of equipment or barbell to the deployment site on the actual Moon?
\end{itemize}

Improving the Apollo 12 concept of operations:
\begin{itemize}
    \item Could you speculate about how to improve this concept of operations on the actual Moon?\\
\end{itemize}

\end{document}